\documentclass[11pt]{article}
\usepackage[top=1in,left=1in,right=1in,bottom=1in]{geometry} %

\usepackage[style=nature]{biblatex}
\usepackage{lscape}
\usepackage{setspace}
\usepackage{graphicx}       %
\usepackage{adjustbox}
\usepackage{booktabs} %
\usepackage{longtable}
\usepackage{tabularx}
\usepackage{threeparttable} %
\usepackage{threeparttablex}
\usepackage{bbding}
\usepackage{amsmath}
\usepackage{ccaption}

\usepackage[labelfont=bf]{caption}
\usepackage{float}

\newcounter{magicrownumbers}

\usepackage{nameref}
\usepackage{tikz}

\usepackage{xcolor}
\definecolor{darkred}{RGB}{138, 25, 17}
\definecolor{darkblue}{RGB}{23, 46, 163}
\definecolor{darkgreen}{RGB}{65, 128, 37}

\usepackage{indentfirst}
\usepackage{times}
\usepackage{authblk}
\usepackage{hyperref} %

\newcommand{\RNum}[1]{\uppercase\expandafter{\romannumeral #1\relax}}

\usepackage{csquotes}       %
\usepackage{makecell}

\hypersetup{
    colorlinks=true,        %
    linkcolor=darkblue,     %
    citecolor=darkred,      %
    urlcolor=darkgreen,     %
    bookmarksopen=true,     %
    bookmarksnumbered=true, %
    pdfstartview=FitH,      %
    pdfpagemode=UseOutlines %
}

\begin{document}

\title{\textbf{The survival of scientific stylization}}

\author[1]{Yuanyuan Shu}
\author[1,*]{Tianxing Pan}
\affil[1]{School of Information Management, Nanjing University}
\affil[*]{Corresponding author. E-mail: pantianxing97@163.com}
\date{}

\maketitle

\begin{abstract}
This study elaborates a text-based metric to quantify the unique position of stylized scientific research, characterized by its innovative integration of diverse knowledge components and potential to pivot established scientific paradigms. Our analysis reveals a concerning decline in stylized research, highlighted by its comparative undervaluation in terms of citation counts and protracted peer-review duration. Despite facing these challenges, the disruptive potential of stylized research remains robust, consistently introducing groundbreaking questions and theories. This paper posits that substantive reforms are necessary to incentivize and recognize the value of stylized research, including optimizations to the peer-review process and the criteria for evaluating scientific impact. Embracing these changes may be imperative to halt the downturn in stylized research and ensure enduring scholarly exploration in endless frontiers.

\end{abstract}

\clearpage
\section*{Introduction}

Stylization, characterized by its role in differentiation and establishing distance from peer works, is extensively explored within cultural spheres such as film, music and artwork~\cite{godart2019explaining,lee2020dissecting,liu2021understanding,negro2022s,sgourev2023relations,askin2017makes}. An idiosyncratic style can endow preeminent creators with a strategic edge~\cite{aghion2005competition,askin2017makes}, yet it simultaneously exposes them to potential risks and failure~\cite{lee2022escaping}. This dichotomy poses a paradox at both the individual and organization level; nevertheless, there exists a prevailing consensus on the significance of diversity within the cultural sector. Contributions imbued with stylistic innovation are indispensable for the flourishing of the cultural milieu as a whole.

Yet, within the purview of scientific exploration, systematized investigation on stylization remains conspicuously sparse, even though such phenomena prevails in spectrum of human creative endeavors. A potential reason for this gap may be that the existing instruments we use to track scientific progress are insufficient. Traditional metrics of research impact, such as citation count, while providing a quantitative measure, often fail to capture the holistic evolution of knowledge within academic fields~\cite{arthur2007structure,funk2017dynamic}.%

To address this, researchers have turned to network-based metrics~\cite{uzzi2013atypical,funk2017dynamic,wu2019large,park2023papers,chu2021slowed} that offer a more sophisticated analysis of scientific development by examining changes within citation networks.
Among these, ``disruption'' metric~\cite{park2023papers} stands out, meticulously charting alterations in citation network to quantify the impact on pre-existing knowledge structures. This retrospective gaze highlights the perpetual forward march of science, wherein today’s cutting-edge discoveries lay the groundwork for tomorrow’s fundamental principles. 

Building upon this analytical shift, we have constructed a text-based metric, termed ``stylization'', designed to quantify the extent to which a scholarly work deviates from established paradigms. This measure relies on intrinsic textual information rather than using the delayed citation dynamic as an indirect proxy for a work's influence within the scientific canon. Our findings, suggest that highly stylized works establish broad connections yet provoke diverse academic critiques, potentially echoing narratives of scientific stagnation~\cite{bloom2020ideas,bhattacharya2020stagnation,chu2021slowed,park2023papers,cui2022aging}.%

Hypotheses ranging from the ``burden of knowledge''~\cite{jones2009burden} to the metaphorical harvesting of ``low-hanging fruit''~\cite{cowen2011great} have been advanced to rationalize this phenomenon. However, in the interstice between the declining visibility of stylized scientific work and the broader deceleration of scientific progress may lay an underappreciated nexus. The momentum of scientific progress hinges upon the cadre of adept scientists, or more pointedly, the vigor of their exploration~\cite{bush1945endless}. Yet, we have observed that diminishing returns in scientific research may be pressuring researchers to focus on outcomes pragmatically rather than exploring their intellectual interests. Substantial, albeit disparate, evidence highlights that the competition for priority may negatively impact the research quality, exerts on scientists’ choices and performance~\cite{hill2021race,bobtcheff2017researcher,tiokhin2021competition,tiokhin2019competition}. This pragmatic shift, while seemingly efficient to the swift response observed during crises~\cite{hill2021adaptability}, might inadvertently stifle the pursuit of pioneering projects with personal styles, potentially contributing to the overall slowdown in scientific advancement.

\section*{Results}

\subsection*{Decline of stylized research over time}

\begin{figure}[H]
    \centering
    \begin{minipage}{0.49\textwidth}
    \center{\includegraphics[width=1\linewidth]{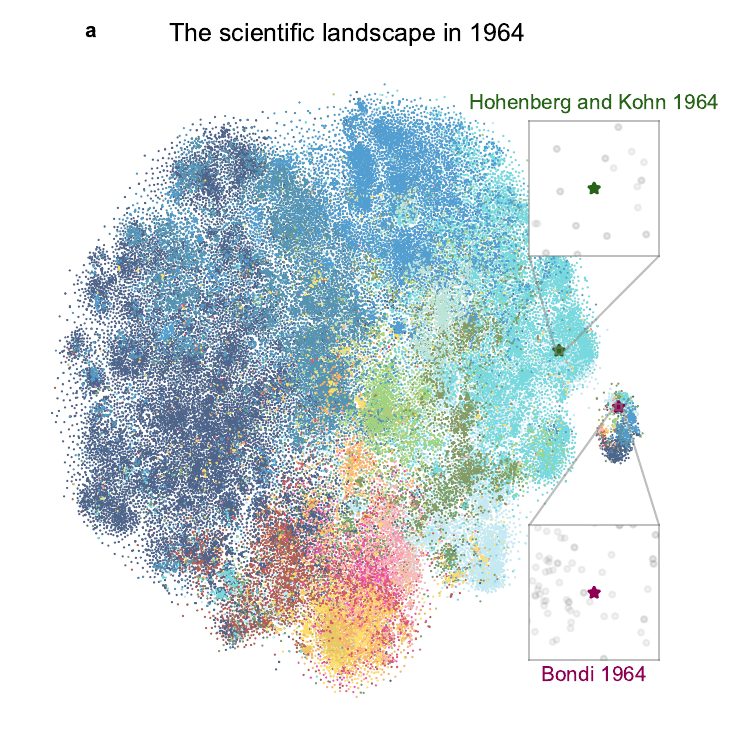}}
    \end{minipage}
    \begin{minipage}{0.49\textwidth}
        \begin{minipage}[h]{1.0\linewidth}
        \center{
        \includegraphics[scale=0.5]{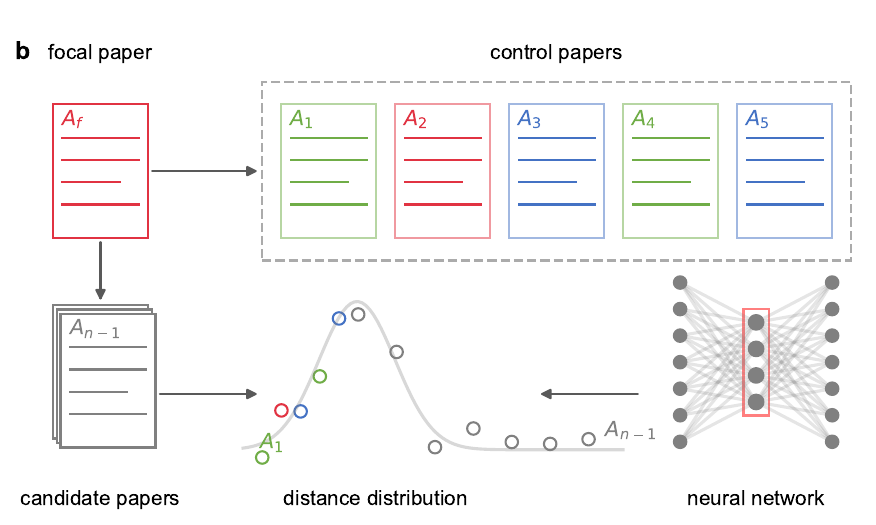}
        }\\
        \end{minipage}
        \vfill
        \begin{minipage}[h]{1.0\linewidth}
        \center{
        \includegraphics[scale=0.5]{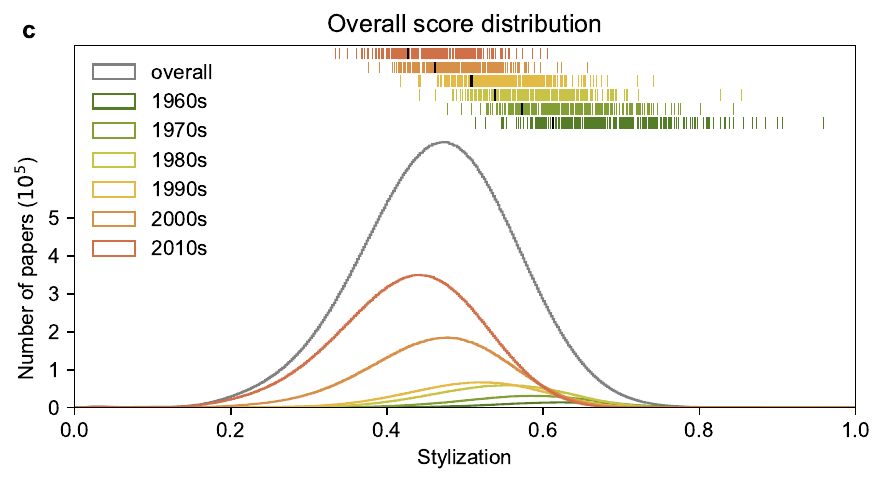}
        } \\
        \end{minipage}
    \end{minipage}
    \caption{
    \textbf{Erosion of scientific stylization.}
    \textbf{a}, This map portrays a 2D projection of a higher-dimensional space, where each point corresponds to a paper published in 1964. The colors denote the respective disciplines, aligned with those shown in Fig.~\ref{fig:pubmed_coverage}. We have employed the \textit{t-Distributed Stochastic Neighbor Embedding (t-SNE)} technique~\cite{van2008visualizing} to reduce high-dimensional vectors for visualization. It is crucial to acknowledge that the positioning of these papers in this 2D space does not precisely mirror their arrangement in the higher-dimensional space. Nonetheless, it serves as an illustrative example, providing an approximate representation of the focal paper's proximity to the entire scholarly community. In the insets on the right, we have zoomed in on two regions of the science map to closely examine the distance between focal paper (colored star) and contemporaneous studies (grey dots).
    \textbf{b}, In the context of a paper published in year $t$ and within field $f$, we have calculated the distance between the focal paper and all other papers from the same year $t$ and field $f$. The five papers that are positioned furthest to the left on the stylized score curve represent the most analogous counterparts. The average of these counterparts collectively represents the level of stylization exhibited by each individual paper.
    \textbf{c}, We have categorized the distribution of stylized metrics into chronological groups, each colored curve here depicts the trend of paper quantity associated with stylization for each decade. To further elucidate disparities across different domains, we have categorized 58,358,209 papers into various fields using the 292 level-1 Field of Study (FoS) tags from the \textit{MAG} dataset. Correspondingly, each colored vertical line on the top represents a specific field, while the black vertical line indicates the average value for the decade. 
    }
    \label{fig:uscore_dist}
\end{figure}

How does the stylized research manifest within the scientific research landscape, and what are the observed trends over time? Addressing this inquiry is pivotal for a nuanced comprehension of the scientific revolution and evolving innovation paradigms. We have commenced this investigation by quantifying the stylization of a paper through neural network embedding, inspired by the prior research~\cite{negro2022s,guzman2023measuring,aceves2023mobilizing}. This approach capitalizes on the observation that highly stylized papers often stand out due to significant disparities compared to their counterparts, as indicated by their distinctive positioning within the scientific knowledge landscape. Specifically, we spotlighted the scientific landscape of 1964, as depicted in Fig.~\ref{fig:uscore_dist}a, highlighting how stylized research distinguishes itself from the contemporary publications. For instance, the green star represents Hohenberg and Kohn's work on \textit{density functional theory}~\cite{hohenberg1964inhomogeneous}, published in the prominent physics journal \textit{Physical Review}. Importantly, this ground-breaking research ultimately led to Kohn being awarded the Nobel Prize in Chemistry in 1998. Meanwhile, the red star in the lower part symbolizes Bondi's paper on \textit{van der Waals radius}, also published in 1964~\cite{bondi1964van}. This paper found practical applications in the \textit{molecular dynamics simulations} that were already in development at the time. Both of these two contributions markedly propelled scientific advancement~\cite{hohenberg1964inhomogeneous,bondi1964van}. 
Indeed, as shown in Fig.~\ref{fig:uscore_dist}a, the former contribution occupies a more distinctive position which is closely linked to related works. Upon closer examination in the magnified inset, Bondi’s paper~\cite{bondi1964van} is nestled within a more competitive milieu, positioning it as a pioneering innovation closely intertwined with mainstream research. In contrast, Hohenberg and Kohn’s work~\cite{hohenberg1964inhomogeneous} remained relatively isolated from the contemporary research, maintaining a more pronounced level of divergence. Therefore, in our study, Hohenberg and Kohn's work is acknowledged for its stylized nature. To quantify this phenomenon, we have measured stylization via the distance between the focal paper and its five most similar counterparts. This metric is pivotal to our analytical framework. Further, we categorized works as ``stylized pieces'' if their measured distance exceeds the mean level within the cohort, while papers falling below this level are deemed more mainstream or popular. This delineation is crucial as it carries significant implications. Specifically, a shorter measured distance suggests a positive engagement with the contemporary research focus and indicates alignment with prevailing research trends. To illustrate the practical application of the stylization concept, we presented in Fig.~\ref{fig:uscore_dist}b the process for computing the stylization score for paper $i$. This measure hinges on the distance of each article from the most similar counterparts in the same field within the same year. And for its robustness checking, our alternative measures of stylization involves computing the average distance from the 10 most similar papers (see Fig.~\ref{fig:uscore_dist_supp}). %

Additionally, we employed the proposed indicator to the Microsoft Academic Graph (MAG) dataset consisting of 58,358,209 papers. It enables us to explore the evolution of stylization over time and across fields since 1960. As depicted in Fig.~\ref{fig:uscore_dist}c, the distribution of stylization score, categorized by decade, exhibits a noticeable leftward shift, with the average stylization decreasing from 0.613 in 1960s to 0.427 in 2010s. Moreover, we dissected this phenomenon further in the uppermost subset of Fig.~\ref{fig:uscore_dist}c, which includes 292 distinct fields spanning decades. This emphasizes the ubiquitous nature of stylization trends across numerous disciplines. Among these, \textit{atomic physics} is exhibited the slowest decline, decreasing from 0.553 to 0.475 over a 50-year period. Conversely, \textit{surgery} have witnessed a rapid decline in stylization from 0.570 to 0.350 as the field expanding. Upon verification, using alternative indicators and selective samples, we have discovered that this decline permeates author cohorts, paper quality, and even scientific communities (see Section~\ref{app:robust_fading}).

\subsection*{Stylized research combined distant components and pioneered new streams}

While existing researches, as exemplified by~\cite{hofstra2020diversity,graves2022inequality}, have rightly underscored the contributions of minorities, such as those pertaining to race~\cite{kozlowski2022intersectional,liu2023non} and gender~\cite{bagues2017does,huang2020historical,ross2022women}, our study posits that exploratory stylized research contributions warrant equal consideration. Thus, we have performed experiments to shed light on the implications of stylized research from two perspectives: knowledge composition and network structure. Here, we have commenced by presenting two illustrative examples, each featuring ten FoS tags. An equal number of combinations facilitates a straightforward comparison of their performance in linking distant knowledge. As Fig.~\ref{fig:uscore_benefit}a depicted, we took into account the study by Folstein et al.~\cite{folstein1975mini} published in 1975 on the \textit{Journal of Psychiatric Research}. This study introduced a dementia assessment questionnaire and exhibited a stylization level that exceeded contemporaneous research in the same field by 0.58 standard deviations. It brought forth 39 novel combinations, with the \textit{Psychiatry} and \textit{Cognitive training} combination proving to be the most influential, garnering 2949 subsequent uptake. As a comparison, Fornell's publication~\cite{fornell1981evaluating} in 1981 on the \textit{Journal of Marketing Research} garnered a comparable level of scholarly recognition, approximately one hundred thousand citations. However, interestingly, its stylization score trailed behind contemporaneous papers by 0.65 standard deviations (see Fig.~\ref{fig:uscore_benefit}b). In terms of knowledge structure, it primarily linked the topics of \textit{Error Analysis} and \textit{Structural Equation Modeling}. This combination has resulted in subtly resonates in subsequent scholarly discussions. Besides, these connections can be seen as a natural extension of the topic \textit{Structural Equation Modeling} methodology in the social sciences during that era, characterized by short knowledge distance between them, measuring only 0.18. In other words, stylized research facilitates the formation of remote combinations of knowledge. 

To this end, we found that these stylized studies bring forth more new knowledge combinations (see Fig.~\ref{fig:uscore_benefit}c-d). In these cases, the average number of new combinations in stylized articles is 4.12, which is 1.15 times that of popularized research. These knowledge combinations display a reduced average frequency of uptake, approximately 0.82 times. Yet, as prior research has noted~\cite{hofstra2020diversity}, less uptake should not be misconstrued as an indicator of diminished value. We further observed that stylized papers increase the probability of remote combinations by 1.26 times and result in 1.35 times more remote combinations. The discrepancy persists, even when accounting for articles that did not introduce novel combinations. Furthermore, measurements based on the average distance of new combinations, 1.14 times, imply that the inclusion of more remote linkages in stylized articles cannot be exclusively attributed to the introduction of additional combinations.

In the aforementioned analysis, stylized works are recognized for their ability to establish connections with more distant knowledge, and over time, the gap between them and popularized papers widens. This persistent trend of bridging distant knowledge enhances the capacity of stylized works to generate novel combinations (see Fig.~\ref{fig:uscore_benefit}c-d). However, an essential query arising from this observed pattern ponders whether stylized research indeed exerts an influence on scientific innovation.

In order to probe into the impact of stylized papers on citation dynamics within the network structure, we have adopted a decomposition approach, drawing inspiration from the work of~\cite{chen2021destabilization}. Contrary to previous measurements~\cite{funk2017dynamic,wu2019large,chu2021slowed,park2023papers,xu2022flat} that amalgamate all references as a unified knowledge base, it is imperative to recognize that certain works often contribute to multiple research topics. In our study, we have engaged an approach that involves an examination of subsequent citations to each precursor work. This method enables us to discern how a scientific work disrupts or pioneers one field while simultaneously consolidating or even strengthening others. This renders consolidation and disruption do not form a dichotomy. Additionally, this approach was formulated based on the existing $CD$ metric, a measurement that was initially introduced by~\cite{funk2017dynamic} and verified by~\cite{wu2019large}. In our study, this decomposition splits the metric into two distinct components: disruption ($D^{\prime}$) and consolidation ($C^{\prime}$). Our analysis of 12,875,121 papers, each receiving at least one citation, unveils that stylized works manifest a diverse array of attitudes towards their knowledge foundation. Importantly, a pronounced divergence among stylized papers when it comes to their potential to disrupt existing researches (Fig.~\ref{fig:uscore_benefit}e). Besides, this divergence is further amplified concerning the process of consolidating existing knowledge (Fig.~\ref{fig:uscore_benefit}f). In essence, these remarkable differences suggest that stylized works fulfill a distinct role compared to popularized works. They alter the citation network structure and have the potential to substantially influence the science stream by establishing connections between distant knowledge domains.

Furthermore, we have conducted word extraction and lemmatization from the article titles in Fig.~\ref{fig:uscore_benefit}g. The outcomes imply that words more frequently appearing in stylized papers are typically linked to the initiation of novel inquiries, articulation of fresh ideas, and exploration of fundamental mechanisms. Conversely, words often associated with popularized research tend to focus on current trending topics like COVID-19, or engage in discussions and consolidations of pre-existing studies. In summary, our findings indicate that stylized research, despite being rooted in existing knowledge, does not deter scientists from carving out unique intellectual paths. These endeavors occupy a unique position within the knowledge space, marked by tenuous connections to existing knowledge and emerging links between remote knowledge domains. They concurrently bolster established communities while presenting challenges to others. This is evident in the increased variability observed in $CD$-type indicators.

\begin{figure}[H]
\centering
\includegraphics[scale=0.5]{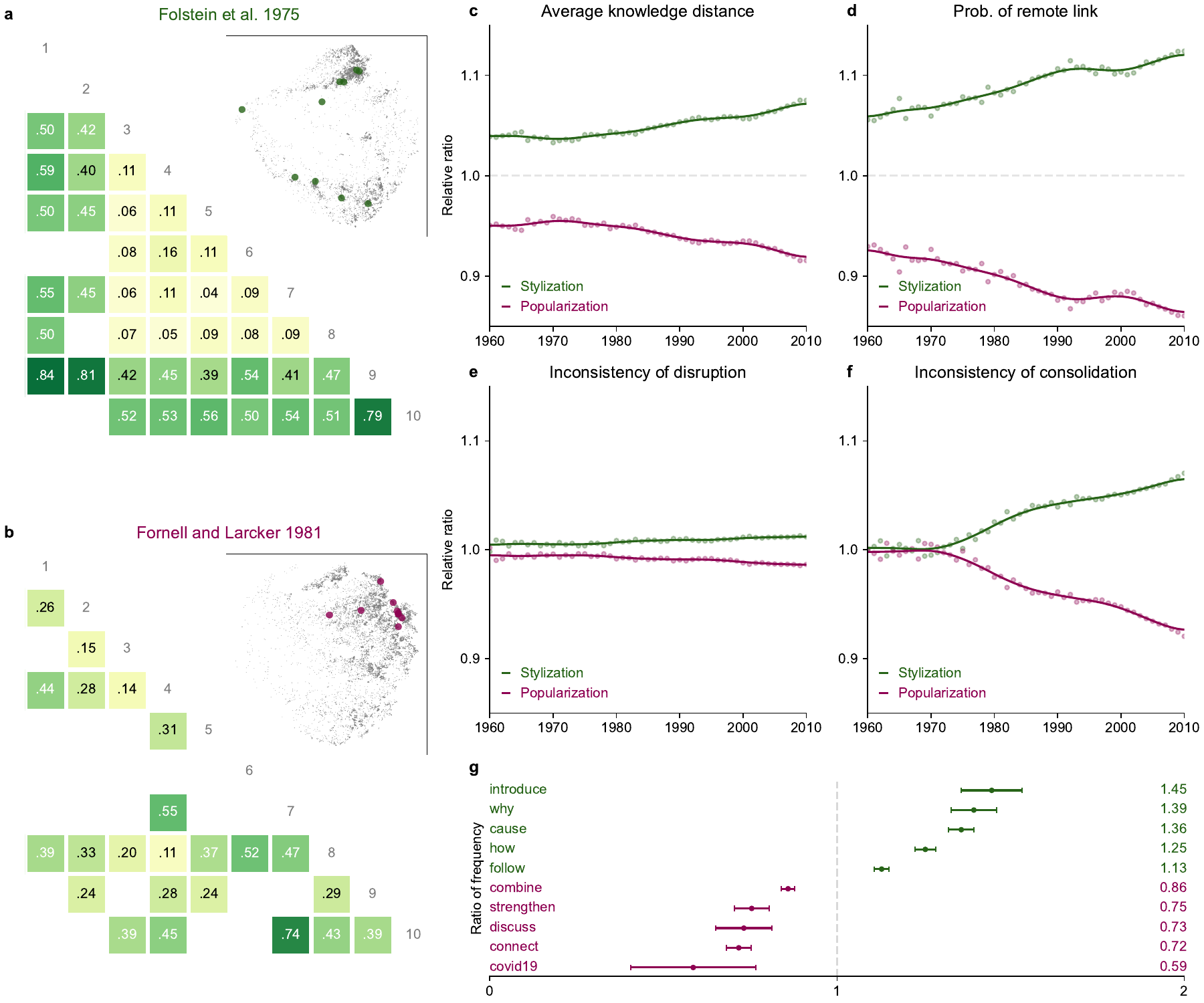} %
\caption{
\textbf{Stylized research facilitates connections among remote knowledge and plays dual role in advancing science.}
(Continued on the following page.)
}
\label{fig:uscore_benefit}
\end{figure}
\begin{figure}[H]
\contcaption{
\textbf{a}, \textbf{b}, We have highlighted two illustrative examples: Folstein 1975 (\textbf{a}) and Fornell 1981 (\textbf{b}). These examples demonstrate the contrasting characteristics of stylized research that differ from popularized research. Stylized papers like Folstein 1975 (\textbf{a}) are shown to connect with more distant knowledge, while Fornell 1981 (\textbf{b}) focuses on concepts that are closer in the knowledge space.
In \textbf{a}, \textbf{b}, for heatmap in the lower left corner, the value in cell represents the distance between knowledge combinations. We have labeled only the new knowledge combinations. For network in the upper right corner, each node in these networks represents a FoS label used within the past 5 years. Colored dots indicate the labels employed by the focal paper, whereas grey dots represent the labels used by other papers within the same field as the focal paper. Notably, since these two maps represent knowledge embeddings from different years, direct point-to-point correspondence between the figures may not exist.
\textbf{c}, \textbf{d}, We have categorized 15.1\% of concept pairs with distances exceeding 0.5 as distant links. Remarkably, stylized papers exhibit a proclivity for uptaking connections with more remote knowledge components, and this inclination becomes increasingly pronounced over time. In 1960, the average knowledge distance ratio between stylized and popularized research stood at 1.04 and 0.95, respectively. However, by 2010, the distance ratio for stylized research escalated to nearly 1.08, while the distance ratio for popularized research dwindled to 0.92 (\textbf{c}). This trend is notably exemplified by the probability of remote links, as illustrated in (\textbf{d}). In 1960, the probability ratio for stylized research was approximately 1.06, whereas for popularized research, it was measured around 0.93. In the contemporary context, the probability ratio for stylized research exceeded 1.12, while that for popularized research declined below 0.9, approaching 0.86. Our observation demonstrated the increasing adeptness of stylized research in establishing distant links within the knowledge space as time progresses.
\textbf{e}, \textbf{f}, We have divided the metric $CD$ into its consolidation (\textbf{e}) and disruption (\textbf{f}) components to explore the dual role that stylized papers play in advancing scientific progress. %
In \textbf{c}-\textbf{f}, We also have performed a Gaussian kernel to fit these scatters into a smoothing curve, illustrating the trends.
\textbf{g}, Upon tokening and lemmatizing with the Python spaCy library, we have discovered that words commonly appearing in stylized papers often revolved around posing queries (``introducte''), floating ideas (``why''), or exploring mechanisms (``cause''). On the other hand, words more likely to appear in popularized research are related to focusing on trending topics (``Covid-19'') or engaging in discussions and consolidation of existing studies (such as ``combine'', ``strengthen'', ``discuss'', ``connect''). Interestingly, certain words previously considered indicative of developmental research (such as ``follow'', ``combine'', ``strengthen'') occupy a relatively intermediate position.
}%
\end{figure}

\subsection*{Stylized papers underwent extended review and received fewer citations}

As proven above, the innovative potential of stylized research manifests in its capacity to chart new pathways and establish distant connections within the scientific landscape. Interestingly, a more compelling question lies in the decline of scientific stylization. As science transforms into a profession, echoing Max Weber's thought~\cite{weber1946science}, it becomes plausible to speculate that researchers start to align their directions and contributions with the expected rewards\cite{merton1968matthew}. This shift in identity necessitates an exploration into potential impediments that stylized research may encounter in achieving commensurate returns.

\begin{figure}[H]
\centering
\includegraphics[scale=0.40]{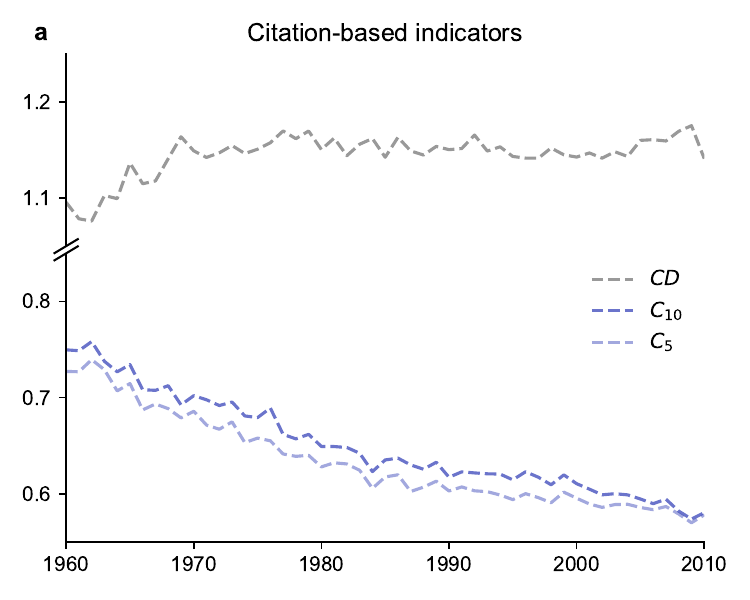}
\includegraphics[scale=0.42]{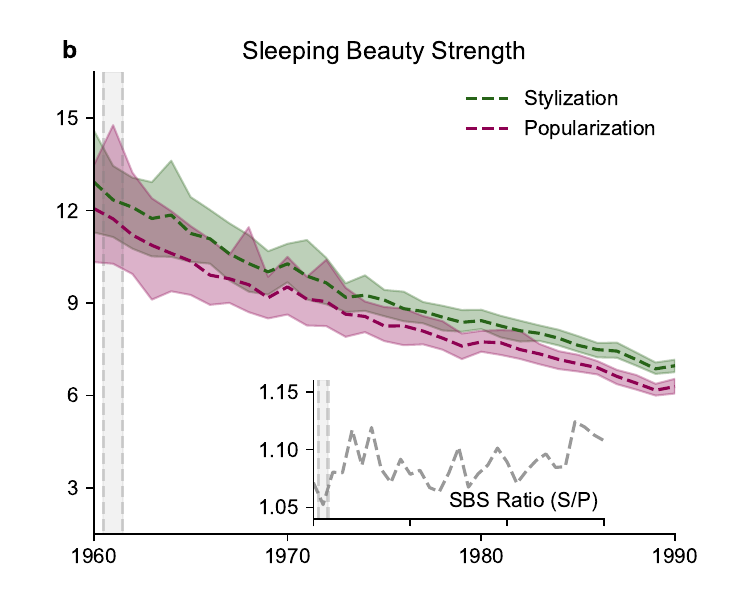}
\includegraphics[scale=0.42]{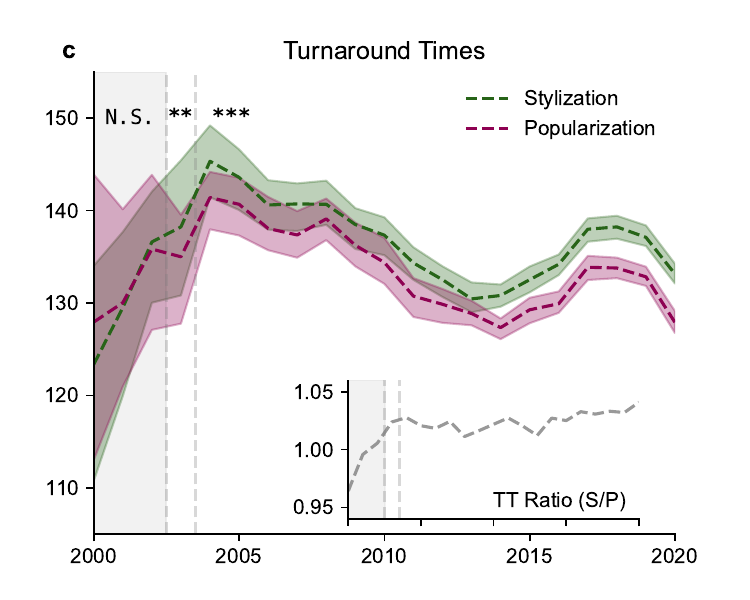}
\caption{
\textbf{Stylized papers receive less attention, fall into a deeper hibernation and undergo longer turnaround time.}
\textbf{a}, Stylized works are subject to citation bias not only in the short term. We conducted a comparative analysis of citation counts within 5-year and 10-year time windows to evaluate the citation performance of stylized and popularized papers. In 1980, popularized papers received 50\% more citations within 5 years, signaling their first advantage in citation. However, this advantage gradually diminished over time. In contrast, stylized papers received recognition over extended windows but were less likely to experience rediscovery. Besides, to assess the innovation quality of stylized papers, we used disruption as a proxy measure, ranging from -1 to 1. With the annual median disruption serving as a cutoff point, we calculated the likelihood of both stylized and popularized papers evolving into disruptive works and computed their ratio. Upon further examination, stylized papers exhibit a higher likelihood of demonstrating disruption in their research.
\textbf{b}, We conducted an analysis of the year-over-year difference in sleeping beauty strength between stylized and popularized papers, with a specific focus on papers published before 1990. This subset of papers has undergone an observation period exceeding 30 years, which enables a more precise evaluation of whether the paper has experienced a resurgence.
\textbf{c}, The turnaround time, defined as the period between the submission date and the acceptance date, emerges as a focus point. Our analysis concerning truncation time predominantly draws from papers published in the span of 2000 to 2020. This selection is guided by the availability of submission history records within this period, as documented in PubMed. Of these, we analyzed truncation times for a total of 4,275,975 papers, uncovering intriguing trends. Interestingly, The data reveals an upward trend in the relative turnaround time for stylized papers. This development signals a widening disparity in turnaround times when comparing stylized and popularized papers. While stylized papers previously had shorter turnaround times than popularized papers, our analysis showed a reversal, with stylized papers undergoing longer review process.
In \textbf{b}, \textbf{c}, the green line represents the stylized papers and red one indicates the popularized ones. The grey shaded region, enclosed by the vertical dashed line, highlights statistical insignificance. *$p < 0.1$, **$p < 0.05$, ***$p < 0.01$. Apart from a few early years, the difference between stylized and popularized papers is statistically significant and continues to widen.
}
\label{fig:year_citation}
\end{figure}

To test if this is the case, we compared the mean citation received by stylized and popularized papers in Fig.~\ref{fig:year_citation}a. This is because citation, as a potent credit token, manifests in scientists in attaining award and promotion. Remarkably, stylized papers tend to obtain fewer citations (0.60 in 5 years, 0.61 in 10 years) compared to their popularized counterparts. This raises concerns about the perceived value and impact of the stylized research again. To address this, we used disruption as a proxy measure to examine the likelihood of stylized papers being classified as destabilizing innovation. Our cumulative findings suggest that the probability of a stylized paper introducing new questions or engaging in theoretical work has remained relatively stable over time. In other words, the growing bias against citing stylized research may not be attributed to a decline in the quality of these studies, as they have proven to be more effective at challenging the existing research. Once again, we compared the citation received within 5-year window and 10-year window, it becomes evident that stylized papers encounter less citation bias over a longer time window (from 0.73 to 0.75 when 1960). However, this re-recognition of scientific value diminishing over time (from 0.58 to 0.58 when 2010).
Whereupon, we also employed the concept of ``sleeping beauty'' (or ``delayed recognition'')~\cite{ke2015defining} to assess the extent to which stylized works are overlooked. We discovered that the stylized pieces tend to remain in a deeper hibernation. (see Fig.~\ref{fig:year_citation}b). Taken together, as the volume of scientific work increases, scientists may find it challenging to allocate sufficient attention to earlier works and re-conceptualize them~\cite{parolo2015attention}.

While our analysis primarily concentrates on the post-publication treatment of stylized works, it constitutes a significant effort to investigate the hypothesis that pre-publication challenges may have contributed to the decline in stylization. To further investigate the review process of the stylized research, we examined the turnaround time, excluding post-acceptance steps like typesetting and licensing, as depicted in Fig.~\ref{fig:year_citation}c. And our dataset, drawn from the PubMed database, is comprised 34,960,700 papers (see Fig.~\ref{fig:pubmed_coverage}). From this database, we identified 4,738,311 papers that included both submission and acceptance dates, allowing for a precise evaluation of the bias in review process. In particular, we observed that during the early 20th century (pre-2003), both stylized and popularized papers underwent similar review durations, with no statistically significant differences. However, by 2020, editors took 4.40\% longer to process stylized papers. This elongated review duration may be attributed to two primary factors. First, stylized works may present challenges in promptly identifying suitable reviewers without college of peers due to their deviation from conventional research~\cite{bammer2016constitutes}. Second, while reviewers are experts in their respective fields, they also serve as gatekeepers of disciplinary boundaries, potentially exhibiting some resistance to stylized works~\cite{fini2023new}. This analysis implies that stylized pieces not only evince biases following their publication, but also face protracted scrutiny to ascertain their publishability.

\section*{Discussion}

The investigation in this paper has identified a concerning decline in the stylized scientific research. The contribution of this study lies in the development and application of a metric to quantitatively assess the extent to which scientific work deviates from the mainstream research. Our findings suggest that stylized research, characterized by its distinctive approach and capacity to bridge disparate knowledge domains, is increasingly scarce within the scientific landscape.

One salient finding is the apparent undervaluation of stylized research when compared with mainstream studies. This phenomenon may be reflective of an academic milieu that disincentivizes risk-taking and pioneering efforts that deviate from conventional paradigms. The downward trend in the production and recognition of such researches poses a potential threat to scientific innovation, given that these works are instrumental in forging new intellectual paths and facilitating interdisciplinary discourse.

Our analysis further reveals that stylized papers not only receive fewer citations but also face longer review time. This may indicates an inherent bias within the peer-review process against unconventional or groundbreaking studies. Despite these challenges, the disruption of stylized research --- their ability to introduce novel questions or theoretical contributions --- has remained consistent over time, underscoring the enduring potential of these works to advance scientific thoughts.

The implications of this study extend beyond the mere identification of trends in academic publishing. They raise critical questions about the metrics for success and recognition within the scientific community. The reliance on citation counts as a proxy for impact may inadvertently marginalize research that, while less cited, contributes profoundly to the diversification and evolution of scientific knowledge.

In addressing these issues, it is essential to consider systemic changes that might encourage the pursuit and acknowledgement of stylized research. Future investigations should explore the factors that contribute to the diminishing production of such research and seek strategies to foster an academic culture that values diversity in scientific inquiry. Moreover, refinements to the peer-review process may be necessary to ensure a fair evaluation of non-traditional research outputs.

\section*{Materials and methods}

\subsection*{Database of publication metadata}

Our data come from two main sources. The first is \textit{MAG} database~\cite{wang2019review}, which contains more than 20 million papers published on more than 45,000 journals and 60 million citations among these papers. The metadata includes citation records, conservative author disambiguation, and other metadata information such as titles and abstracts. Taking into account the transformation of the scientific landscape and its structure after the post-war period, our analysis focus on the publications from 1960 to 2010. To construct the knowledge base, we also incorporated data from works published prior to 1960. Likewise, we considerd these prior works when calculating metrics like disruption. Also, we used the 19 level-0 FoS tags and 292 level-1 FoS tags provided by MAG as discipline and field information.

\subsection*{Database of publication history}

Second, the publication history data in this study is sourced from the \textit{PubMed} database, where archived 34,960,700 publications primarily cover fields such as \textit{Medicine}, \textit{Biology} and \textit{Chemistry} and so on. Our disclosure of the database coverage , as shown in Fig.~\ref{fig:pubmed_coverage}, also highlights its diversity, encompassing fields such as \textit{Psychology} and \textit{Materials science}. Notably, we excluded 27,215 papers that are shared the same Digital Object Identifier (DOI), an occasional scenario occurring primarily with earlier papers published on the same page. For instance, five papers, whose PubMed reference number (commonly known as PMID) ranging from 16398630 to 16398634, point to the same DOI 10.5694/j.1326-5377.2006.tb00098.x.

As part of our methodology, we applied stringent exclusion criteria, which led to the removal of papers with turnaround times below 30 days. This meticulous selection process aimed to mitigate potential errors arising from submission and acceptance date records. Similarly, we excluded papers with exceptionally prolonged turnaround times, exceeding 1,000 days, although such instances were rare, representing outliers in the review process. The former may arise when publishers submit the acceptance date or revised date as the submission date to both \textit{Crossref} and \textit{PubMed}, while the latter is attributed to the fact that, although many manuscripts might go through even longer review cycles, only a small fraction of them will remain in a journal's publishing system for over three years. Our final dataset, posting these exclusions, encompasses 4,738,311 papers. It is crucial to emphasize that these exclusions are implemented to maintain the validity of our analysis and have no bearing on our overarching conclusions, as substantiated through validation tests (see Fig.~\ref{fig:sbs_lag_supp}).

\subsection*{Construct and validate stylization}
Word embedding techniques stand as a powerful instrument in natural language processing, for understanding the interplay between the structure and property within material science field~\cite{tshitoyan2019unsupervised,vasylenko2021element,pei2023toward}, repurposing drugs for diseases where treatment options remain scarce~\cite{sourati2023accelerating}, aligning semantic meanings across varied linguistic contexts~\cite{thompson2020cultural}, tracing the human trajectory in scientific mobility~\cite{murray2020unsupervised}. Building on these diverse and powerful capabilities, our study employed word embedding techniques to encode each academic paper as a continuous vector $z_n$ in the embedding space. Therein, $n$ represents the dimension of the vector, which is 768. Then, to construct the stylization score, we collected all papers $P_t^f$ published in the same cohort as candidates, for a paper $i$ published on year $t$ in the filed $f$. Specifically, we obtained the corresponding embeddings $Z_t^f$ of all these papers from a model that was trained on the entire corpus. The continuous vector matrix with $\vert{P_t^f}\vert \cdot n$ could represent the semantic information of all papers $P_t^f$ published in the same cohort as candidates. Subsequently, we created a distance matrix $D_t^f$ with $\vert{P_t^f}\vert \cdot \vert{P_t^f}\vert$ based on these vectors. Then we compared each paper with 5 most similar papers and averaged the distance from them as the $stylization$ score. To address the anisotropic challenge associated with word vectors, we engaged in a rotation procedure concerning word vectors $Z_t^f$ within the same cohort to eliminate shared components. Additionally, it is worth noting that the expansion of the scientific frontier not only gives rise to novel disciplines but also nurtures the growth of pre-existing ones~\cite{mu2017all}. To ascertain whether the decline in stylization is a byproduct of the quantity of the most similar papers, we computed stylization based on the proximity of the nearest 5\% of papers within the knowledge space. And robustness analyses confirmed that neither the rotation postprocessing nor the count of articles as candidates exerts any influence on our findings (see Section~\ref{app:vs}).

In addition, we assembled a dataset comprising 10,641 twin pairs of papers from \textit{PubMed}. These papers, specifically chosen without co-authors, are spanned from 1964 to 2018 and covered 118 fields of study. Significantly, among these pairs, 1135 (10.67\%) were published in a back-to-back form. This particular publication format is a deliberate strategy that is employed by journal editors to fairly attribute credit to simultaneous discoveries (see Section~\ref{app:vs} for more details on simultaneous discoveries). To validate the reliability of the stylization measure, we have evaluated whether it allocates similar scores to twin paper pairs, as illustrated in Fig.~\ref{fig:twins}a. 

At the same time, we randomly selected a paper from the same field and year to serve as a ``control paper''. It is expected that the control papers should demonstrate significantly greater differences in stylization scores when compared to the experimental group. As shown in Fig.~\ref{fig:twins}b, the difference in stylization score between twin paper pairs is less than this between one of the pair and a randomly selected paper published in the same year and field. Specifically, 88\% of all twin paper pairs exhibit scores that differ by no more than 0.05, while this threshold is met by fewer than 40\% of the pairs in the randomized control group. Notably, this observation remains consistent even when the stylization score is converted to quantile. Moreover, the Wilcoxon rank sum test corroborates that there is no statistical difference in stylization score between twin paper pairs, but there is a significant difference between the twin papers and random sample. 

To better understand and explain the variations in stylization score among twin papers, we conducted further exploration. We compared the five most similar papers used in the stylization score and found that the majority of twin paper pairs were among each other's top five most similar papers. We also computed the overlap rate of similar neighbors for stylization, as depicted in Fig.~\ref{fig:twins}c. Although no pair of twin papers shared the exact same similar neighbor papers, the majority of twin papers did have some overlapping similar neighboring papers.

Noteworthy is the calculation of stylization, which relies on continuous vectors within a high-dimensional space. This characteristic implies that authors are striving to enhance their work's performance on this metric face challenges when trying to manipulate it through selective journal citations or omission of specific references. Furthermore, it is vital to distinguish this approach from the calculation of the disruption metric, as they differ significantly in two key aspects. Firstly, the stylization metric is text-based metrics that does not require a considerable citation count for the work under scrutiny. Secondly, unlike the disruption metric, it does not require a lengthy observation period. This makes the stylization metric more immediately applicable and less prone to manipulation, thus enhancing its robustness in evaluating the scientific paper. 

\subsection*{Citation normalization across year and field}

The papers included in our analysis are each assigned to at least one of the 292 fields provided by MAG. We counted the number of citations that these papers are received and normalized by year and field. Specifically, we implemented the methodology that are provided in~\cite{radicchi2008universality}. We also calculated the number of citations $C_{5}$ and $C_{10}$ for each paper accrued within 5 and 10 years after publication. Referring to existing works, we used the unscaled $C_{5}$ and $C_{10}$ in the main text and reported the results based on the standardized post-citation in Section~\ref{app:rm}. In the regression analyses, we employed all these measures. This approach allows us to observe and understand the impact of a paper in its respective field while considering the variations across different fields and years.

\pagebreak
\printbibliography

\pagebreak
\section*{Data and materials availability}
Data associated with this research will be available in the public repository at \textit{url}.

\section*{Acknowledgments}
We thank Libo Sheng from Nanjing University for providing the dataset on the links between Lasker awardees and their corresponding papers. All errors are our own.

\section*{Author Contributions}
Yuanyuan Shu performed empirical analyses and drafted the manuscript. Tianxing Pan conceived the project, collected research data, visualized related results and revised the manuscript.

\section*{Competing interests}
The authors declare no competing interests. 

\section*{Additional Information}

\noindent
\textbf{Supplementary Information} is available for this paper.

\noindent
\textbf{Correspondence and request for materials} should be addressed to Tianxing Pan.

\clearpage

\renewcommand\contentsname{Supplementary Information}
{\let\clearpage\relax \tableofcontents}
\thispagestyle{empty}
\begin{refsection}
\setcounter{figure}{0} 
\renewcommand{\thesection}{S\arabic{section}}
\setcounter{table}{0}
\renewcommand{\thetable}{S\arabic{table}}
\setcounter{figure}{0}
\renewcommand{\thefigure}{S\arabic{figure}}

\clearpage
\section{Validating stylization}\label{app:vs}
In this section, we outline the method for identifying twin papers and using them to corroborate the validity of our proposed stylization indicator, which should yield similar stylization scores for twin papers if the indicator is effective.

Remarkable coincidences frequently emerge in the realm of scientific discovery, characterized by multiple scientists or research teams independently achieving significant breakthroughs concurrently. This phenomenon, often referred to as ``multiple discovery''~\cite{merton1957priorities,patinkin1983multiple} or ``simultaneous invention''~\cite{ogburn1922inventions}, stands as a testament not only to the vibrancy and ingenuity within the scientific community but also highlights the objectivity and reproducibility inherent in scientific methodologies~\cite{merton1961singletons}. Notable examples include the independent discoveries of \textit{Sir Isaac Newton} and \textit{Gottfried Wilhelm Leibniz}, who developed the principles of calculus in the late 17th century. Likewise, in the early 2010s, there was the nearly simultaneous revelation of the revolutionary CRISPR-Cas9 gene-editing system, with contributions from both \textit{Jennifer Doudna} and \textit{Emmanuelle Charpentier}~\cite{jinek2012programmable} who were awarded the Nobel Prize, as well as \textit{Feng Zhang}~\cite{cong2013multiplex} who held the patent. Similarly, academic publishers such as PLOS, AMS\footnote{According to ASM's policy framework, manuscript submissions are not subject to rejection based solely on the subsequent publication of a competing paper. This protective stance extends to scenarios where similar findings are published or posted on a preprint server, typically allowing for a grace period of up to six months before the manuscript's formal submission date. Extensions to this grace period may be considered in cases where a preprint of the identical manuscript version was posted during the stipulated timeframe.} and eLife~\cite{marder2017beyond} have implemented \href{https://journals.asm.org/scooping-policy}{Scooping Protection Policy} (sometimes termed \href{https://academic.oup.com/genetics/pages/general-instructions#AboutTheJournal}{No Scoop Policy}) to safeguard the intellectual contributions of authors against preemptive publication by competing researchers. These policies recognize the existence of multiple discovery and accredit twin papers.

\begin{displayquote}
    ... the editors appreciate that competing studies often complement each other, recent publication of similar articles by others does not necessarily preclude consideration of a manuscript for publication in \textit{Genetics}...(\textit{Genetics} is a journal published by Oxford University Press.)
\end{displayquote}

To date, studies on simultaneous discoveries in scientific research, often referred to twin papers, have previously examined aspects such as gender disparities~\cite{bikard2022standing} and geographic distance~\cite{bikard2020bridging} in innovation outcomes. Twin papers serve as a quasi-experiment, enabling precise control over the special variables in similar scientific research contexts. This approach provides valuable insights into the nuances of scientific discovery, contributing to our understanding of how different factors influence the recognition and impact of innovations in science. In our research, these instances of ``multiple discovery'' provide a unique opportunity to validate scientific metrics by examining the papers within each set of twins. Inspired by aforementioned studies, we aimed to assign the similar stylization scores to twin papers, assuming this metric proves effective.

\begin{figure}[H]
\centering
\includegraphics[scale=0.5]{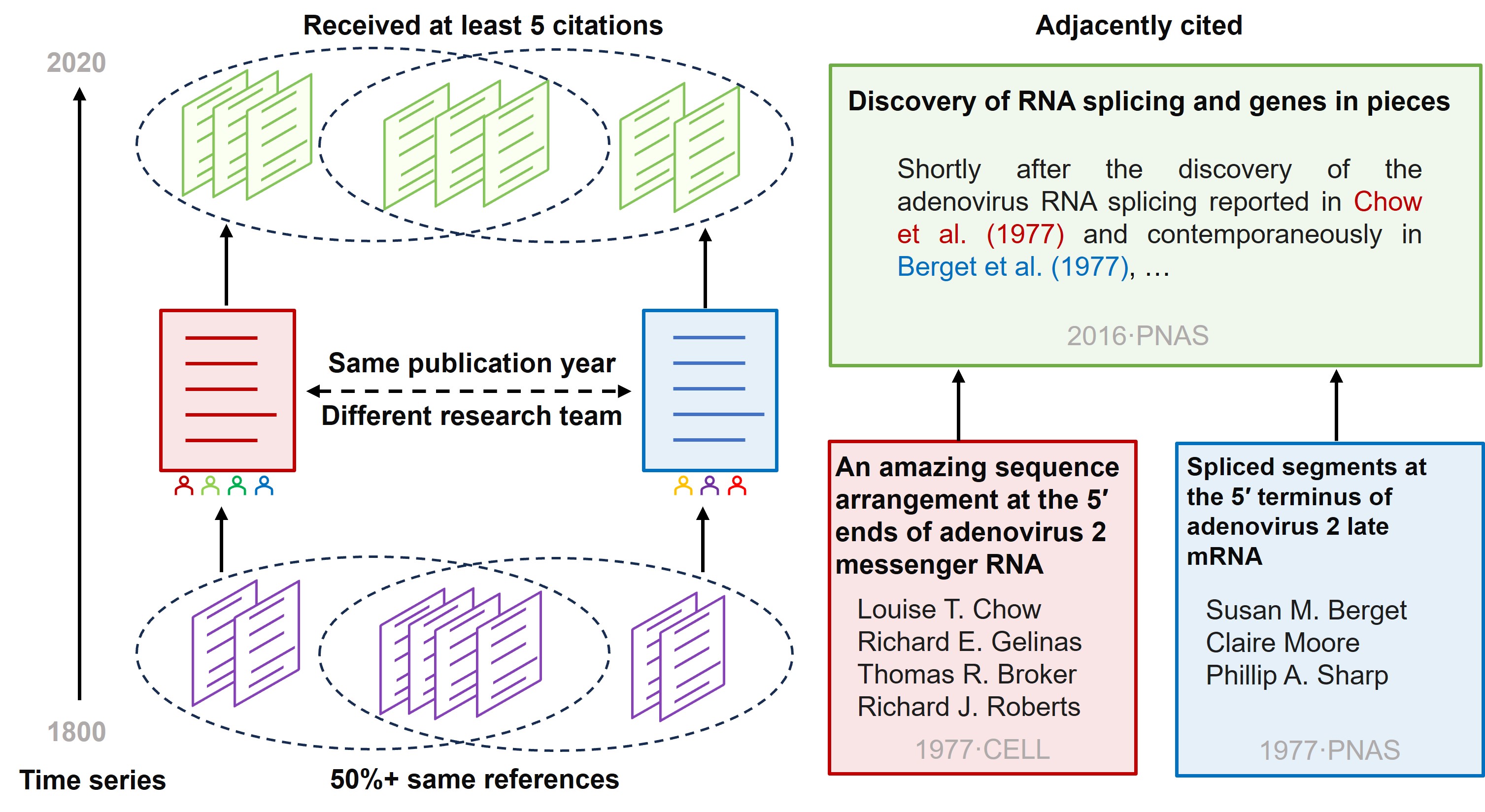}
\caption{
\textbf{Illustration of identifying the twin papers.} 
To further validate the reliability of the stylization score used in the main text, we conducted a series of experiments hinging on the concept of ``twin papers''. Given the diverse characteristics described above and the identification methods employed in existing research~\cite{bikard2022standing,bikard2020bridging,bikard2020idea}, for each cited paper, we established a pool of candidates from the period 1800 to 2020, selected based on shared references, publishing journal, and publication year. %
For example, we traced back from a 2016 PNAS journal on RNA to the 1977 Cell journal and the 1977 PNAS journal, which were cited in the same sentence. We then extracted relevant information such as titles and authorship details from these twin papers, which pertained to the same research terms. This method enabled us to effectively validate the reliability of the stylization score by observing its consistency across similar or  ``twin '' research studies.
}
\label{fig:b2b_example}
\end{figure}

In Fig.~\ref{fig:b2b_example}, we illustrated the comprehensive process of defining twin papers. These twin papers were identified as those cited more than three times within the same parentheses or in adjacent sentences. Specifically, they do not share any of the same authors but have at least 50\% of the same references. In essence, these papers drew inspiration and built their knowledge base from similar literature, often arriving at similar or closely related conclusions, which were frequently discussed within the academic community. To closely align with the concept of \textit{simultaneous discovery}, we further required that these twin papers share the same publication year. This constraint particularly reduces the likelihood of later-published papers merely mimicking their earlier counterparts. In some cases, two papers within twin pairs are published on the same journal even cited each other, especially when editors from top journals chose to publish them back-to-back (an arrangement often seen in scientific journals such as Nature, Science and PNAS). In Table\ref{tab:detailed_twin}, we presented the top 10 co-cited pairs from these twins, along with their metadata and corresponding news. In particular, the distinction of ``different teams'' means that paired output by same researchers, such as~\cite{chetty2022sociala,chetty2022socialb}, will not be recognized as ``twin papers''. The requirement for similar references signifies their reliance on shared literature, which in turn leads to the formulation of comparable conclusions that are commonly discussed within the academic community.

\begin{table}[ht]
\centering
\caption{
    \textbf{Elaborating details about certain twin papers.}
    We performed an exhaustive analysis of the identified twin pairs, narrowing our focus to the top 10 pairs based on co-citation count. %
    Here, we employed \textit{RefSim}, a metric that measures the degree to which these papers share references. This measurement is essential for understanding the extent to which twin papers draw from the same body of literature and the overlap in their knowledge sources. Additionally, we explored the publication status of these twin papers. The term \textit{B2B} signifies whether they were published as back-to-back papers. We also examined cases where these twin papers were released in the same issue, verifying whether there was any related scientific news in that issue. If relevant, we included the titles of these news articles (italic in the column ``Pairs''), offering a broader perspective on the context and impact of the twin papers. Lastly, for clarity, \textit{N Engl J Med} refers to \textit{The New England Journal of Medicine}, while \textit{Nat Genet} stands for \textit{Nature Genetics}.
}
\resizebox{\linewidth}{!}{
    \begin{tabular}{llllll}
    \toprule
     & Pairs & Year & JournalName & B2B & RefSim \\
    \addlinespace
    \midrule
    1 & \makecell[l]{Cold-Activated Brown Adipose Tissue in Healthy Men\\Functional Brown Adipose Tissue in Healthy Adults\\\textit{Brown Adipose Tissue — When It Pays to Be Inefficient}} & 2009 & N Engl J Med & 0 & 0.611 \\
    \addlinespace
    2 & \makecell[l]{Targeting of HIF-$\alpha$ to the von Hippel-Lindau Ubiquitylation Complex by $O_2$-Regulated Prolyl Hydroxylation\\HIF$\alpha$ Targeted for VHL-Mediated Destruction by Proline Hydroxylation: Implications for $O_2$ Sensing\\\textit{---------------}} & 2001 & Science & 1 & 0.532 \\
    \addlinespace
    3 & \makecell[l]{Cloning of the Gene Containing Mutations that Cause PARK8-Linked Parkinson's Disease\\Mutations in LRRK2 Cause Autosomal-Dominant Parkinsonism with Pleomorphic Pathology\\\textit{---------------}} & 2004 & Neuron & 1 & 0.624 \\
    \addlinespace
    4 & \makecell[l]{Safety and Efficacy of Gene Transfer for Leber's Congenital Amaurosis\\Effect of Gene Therapy on Visual Function in Leber's Congenital Amaurosis\\\textit{Preliminary Results of Gene Therapy for Retinal Degeneration}} & 2008 & N Engl J Med & 1 & 0.589 \\
    \addlinespace
    5 & \makecell[l]{Receptor Interacting Protein Kinase-3 Determines Cellular Necrotic Response to TNF-$\alpha$\\Phosphorylation-Driven Assembly of the RIP1-RIP3 Complex Regulates Programmed Necrosis and Virus-Induced Inflammation\\\textit{---------------}} & 2009 & Cell & 1 & 0.519 \\
    \addlinespace
    6 & \makecell[l]{Drosophila odorant receptors are both ligand-gated and cyclic-nucleotide-activated cation channels\\Insect olfactory receptors are heteromeric ligand-gated ion channels\\\textit{Current views on odour receptors}} & 2008 & Nature & 1 & 0.740 \\
    \addlinespace
    7 & \makecell[l]{Targeting Malaria Virulence and Remodeling Proteins to the Host Erythrocyte\\A Host-Targeting Signal in Virulence Proteins Reveals a Secretome in Malarial Infection\\\textit{The Malarial Secretome}} & 2004 & Science & 1 & 0.748 \\
    \addlinespace
    8 & \makecell[l]{A mouse Mecp2-null mutation causes neurological symptoms that mimic Rett syndrome\\Deficiency of methyl-CpG binding protein-2 in CNS neurons results in a Rett-like phenotype in mice\\\textit{---------------}} & 2001 & Nat Genet & 1 & 0.528 \\
    \addlinespace
    9 & \makecell[l]{Evidence that stem cells reside in the adult Drosophila midgut epithelium\\The adult Drosophila posterior midgut is maintained by pluripotent stem cells\\\textit{Editorial Summary}} & 2006 & Nature & 1 & 0.765 \\
    \addlinespace
    10 & \makecell[l]{$Prdm9$ Controls Activation of Mammalian Recombination Hotspots\\PRDM9 Is a Major Determinant of Meiotic Recombination Hotspots in Humans and Mice\\\textit{Homing in on Hotspots}} & 2010 & Science & 1 & 0.560 \\
    \addlinespace
    \bottomrule
    \end{tabular}
}
\label{tab:detailed_twin}
\end{table}

In our discussion, we presented two illustrative examples. The first example involves the research conducted by Laguette et al.~\cite{laguette2011samhd1} and Hrecka et al.~\cite{hrecka2011vpx}, which investigates the relationship between \textit{Vpx}, \textit{SAMHD1}, and \textit{HIV-1}. This work was summarized by Nature in a viewpoint article titled ``Going to the Watchers''~\cite{lim2011going} within the same issue (\textit{Nature}~474,~7353). The study by Laguette et al. was assigned a stylization score of 0.217, whereas the subsequent work garnered a score of 0.197. The relevant sentences from the viewpoint article are presented below.

\begin{displayquote}
In this issue, Laguette et al.~\cite{laguette2011samhd1} and Hrecka et al.~\cite{hrecka2011vpx} identify the host protein targeted by Vpx.
\end{displayquote}

Similarly, another example is from \textit{Science}~292,~5516. They are mentioned in the same comment ``How do cells sense oxygen?''~\cite{zhu2001cells} and even in the same sentence. The former was assigned 0.268, while the latter received a score of 0.305. The related sentences in the news are as follows.

\begin{displayquote}
In their Perspective, Zhu and Bunn discuss new findings (Ivan et al.~\cite{ivan2001hifalpha}, Jaakkola et al.~\cite{jaakkola2001targeting}) that reveal how the \textit{HIF} transcription factor, which switches on a group of hypoxia-response proteins, is itself regulated by changes in oxygen tension.
\end{displayquote}

\begin{figure}[H]
\centering
\includegraphics[scale=0.5]{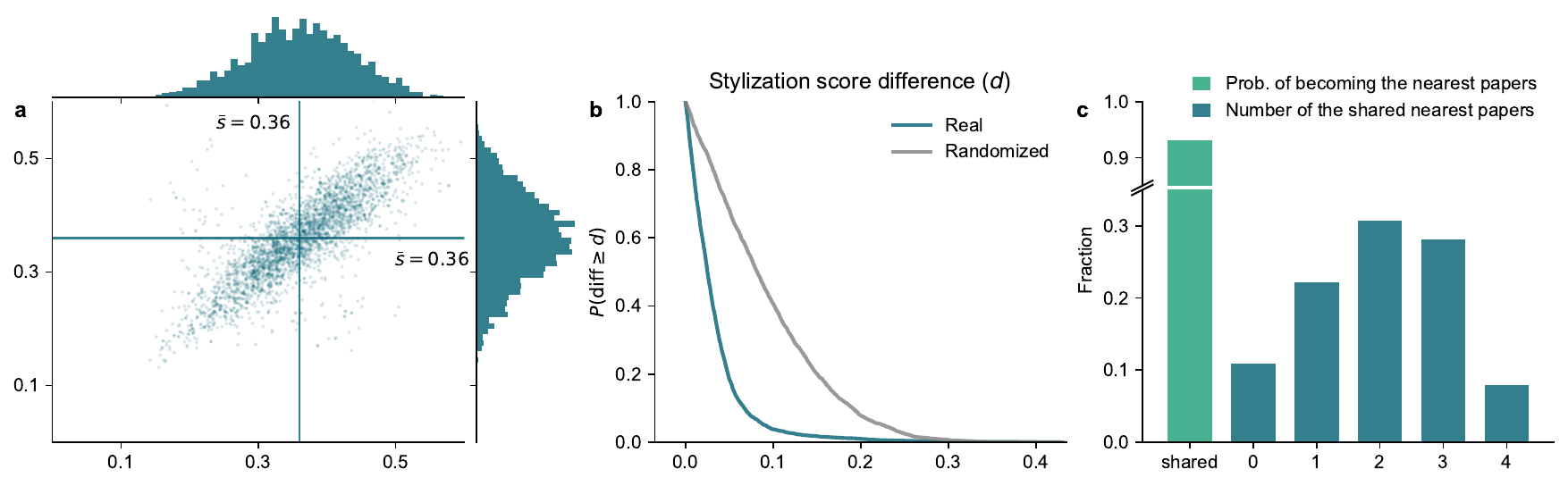}
\caption{
\textbf{Twin papers share similar stylization score.}
\textbf{a}, The stylization scores for two papers in each pair are plotted on the x-axis and y-axis, revealing a positive linear correlation (Pearson correlation coefficient = 0.80). In the upper inset along the x-axis, we presented the distribution of stylization score for the earlier paper in the twin pair, while the right inset along the y-axis corresponds to the score distribution of the later paper. Notably, these twins have mean scores $\overline{S}$ at approximately 0.36, with similar overall distribution. Upon closer examination of the scatter plot, a strong resemblance in the score distribution characteristics for each pair of twins.
\textbf{b}, The survival distribution of the difference in stylization scores is depicted, with the grey curve being computed based on one of the twins and its corresponding randomized control group.
\textbf{c}, In the process of computing stylization, we matched the five most similar papers (neighbours) in each paper (see Fig.~\ref{fig:uscore_dist}b). The first green bar on the left signifies that among more than 90\% of twins, they become each other's 5 most similar papers. Following this, the five cyan bars on the right represent the overlap count of twin papers. The results indicate that no twin papers share exactly the same neighbours, but the majority of twin papers do share some neighbours.
}
\label{fig:twins}
\end{figure}

Notably, the paired papers identified here may not be entirely identical scientific discoveries. However, their primary findings can be inferred to be roughly similar. When combined with the illustrated media coverage and the extraordinary back-to-back publication rates (10.67\%), it becomes evident that these twin discoveries can be considered as similar and simultaneous to a certain extent. As such, they serve as an effective tool for evaluating the stylization metric. The experiments shown in Fig.~\ref{fig:twins} confirm the validity of our proposed indicators, and a detailed interpretation could be found in the Materials and methods section.

\clearpage
\section{Robustness check for stylization fading}\label{app:robust_fading}

In this section, we carry out a series of experiments to investigate the potential decline in scientific stylization as a robustness check. This includes the various measures we used and different samples we investigated. 

\subsection*{Alternative measures}

\begin{figure}[H]
\centering
\includegraphics[scale=0.5]{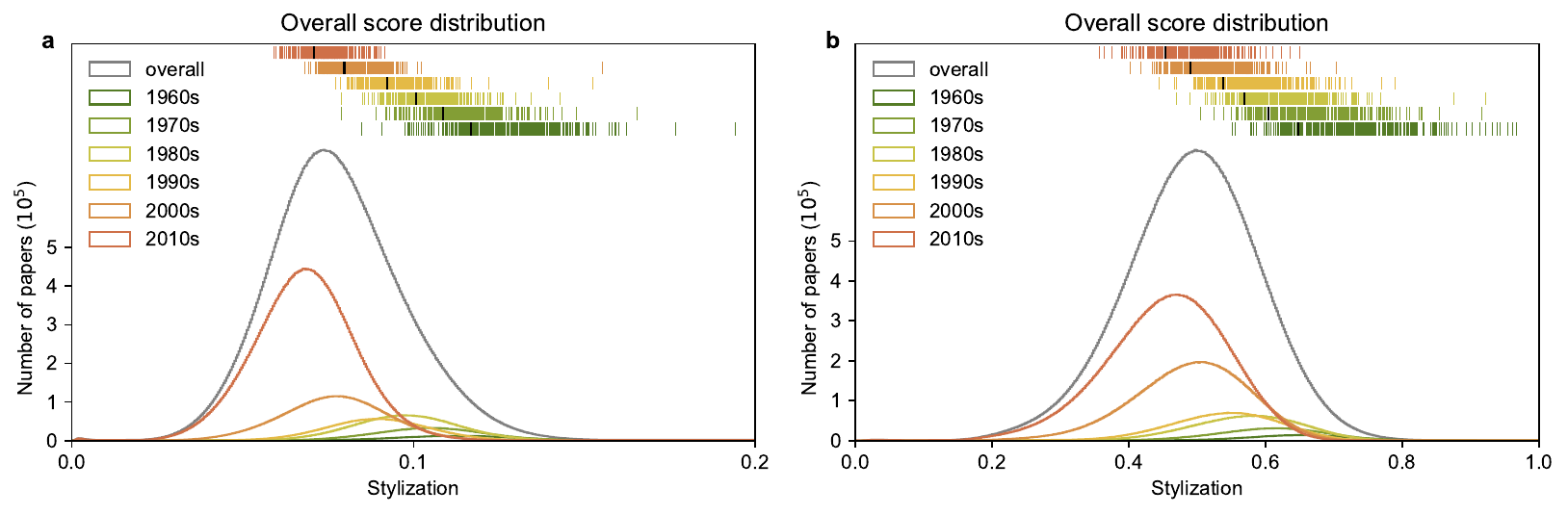}
\caption{
\textbf{Robustness check for alternative specifications.}
We calculated the stylization between focal paper and 5 most similar papers prior to rotation (\textbf{a}), and expanded the number of papers considered for calculating the average distance from 5 to 10 (\textbf{b}). These results strongly indicate that the observed decline in scientific stylization, as elucidated in our study, are not a artefact of our specific approach.
}
\label{fig:uscore_dist_supp}
\end{figure}

While it is recognized that science is multiple independent discovery, accurately identifying the exact number of competitors for a scientific paper remains a challenge. For instance, in some academic disciplines, merely considering the five most recent papers may not afford an exhaustive assessment of a focal paper's distinctive character. Consequently, in Fig.~\ref{fig:uscore_dist_supp}, we have adopted the mean distance with the 10 closest papers as an alternative. Furthermore, given the rapid expansion of numerous scientific fields, researchers may increasingly find themselves in a more competitive landscape, even while they venture into uncharted territories. Simply put, scientists are now more likely to encounter similar counterparts for their work. In light of this, we used percentiles rather than a fixed count to determine the number of candidates considered in the distance calculation. Futhermore, we included the 5\% of papers in the same cohort that are closest in the scientific vectore-space to be included in the calculation. Fig.\ref{fig:uscore_dist_supp}, based on these measures, exhibits a similar downward trend as the results reported in the main text (see Fig.~\ref{fig:uscore_dist}c). This consistent pattern, irrespective of the methods, reinforces the robustness of related findings.

\subsection*{Selective samples}

\begin{figure}[H]
\centering
\includegraphics[scale=0.5]{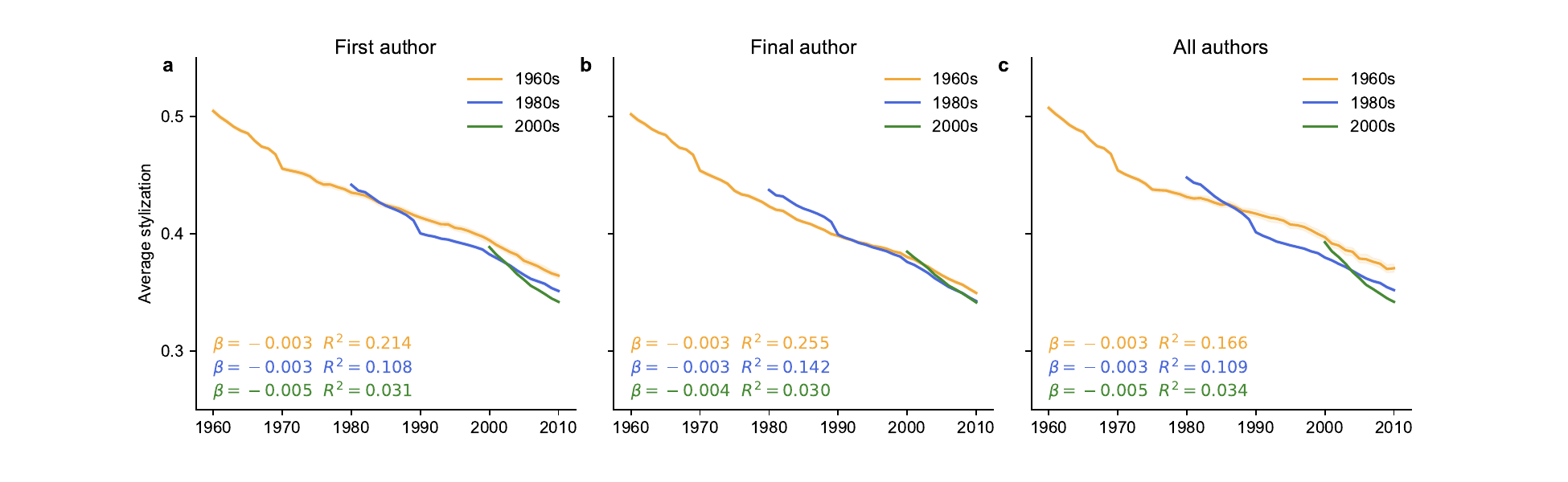}
\caption{
\textbf{Stylization of different cohorts over time.}
We grouped scientists based on the year they entered academia and observed that the papers they authored as the first author (\textbf{a}), the last author (\textbf{b}) or all authors (\textbf{c}) also are exhibited a decline in stylization over time.
In \textbf{a}-\textbf{c}, the solid line depicts the evolution of the mean stylization score from 1960 to 2010, with the shaded area representing the 95\% confidence interval. Additionally, in the lower-left corner, we provided the slope ($\beta$) and the coefficient of determination ($R^2$) based on linear regression estimates. 
}
\label{fig:cohort_decline}
\end{figure}

\begin{figure}[H]
\centering
\includegraphics[scale=0.5]{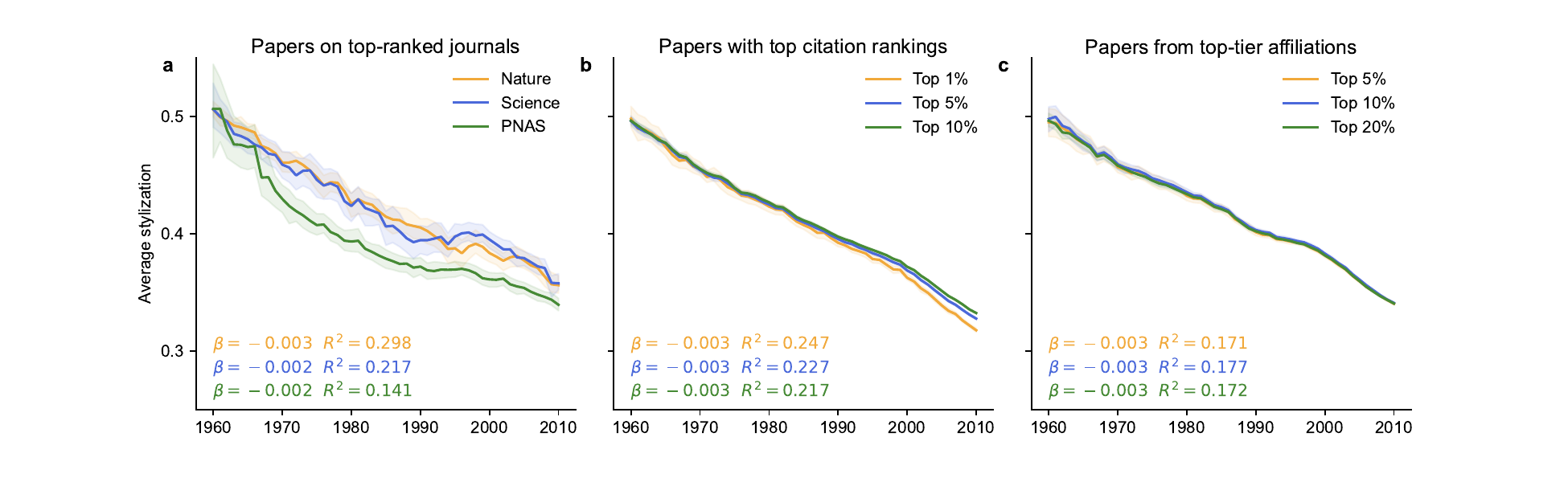}
\caption{
\textbf{High quality papers are losing their stylization.}
\textbf{a}, There is also a decline in stylization for papers published in top-tier journals, such as Nature, Science, and PNAS.
\textbf{b}, We examined papers in the top 1\%, 5\%, and 10\% citation rankings and identified similar trends.
\textbf{c}, We screened the top 5\%, 10\%, and 20\% of affiliations, and their rankings were determined based on the citation graph using the PageRank algorithm. In \textbf{a}-\textbf{c}, the solid line depicts the evolution of the mean stylization score from 1960 to 2010, with the shaded area denoting the 95\% confidence interval. Moreover, in the lower left corner, we presented the slope ($\beta$) and the coefficient of determination ($R^2$) derived from linear regression estimates. 
}
\label{fig:quality_decline}
\end{figure}

In Fig.~\ref{fig:cohort_decline}, we categorized scientists into different cohorts by their academic entry year and grouped papers based on cohorts regardless of their authorship position. We observed a decline in stylization over time in the papers they authored. Furthermore, as depicted in Fig.~\ref{fig:quality_decline}, our analysis focus on papers published in the top-ranked journals, highly cited papers (1\%, 5\%, 10\%). And building upon the existing research~\cite{jones2008multi}, we defined three different tiers of top affiliations (5\%, 10\%, 20\%).

\begin{figure}[H]
\centering
\includegraphics[scale=0.5]{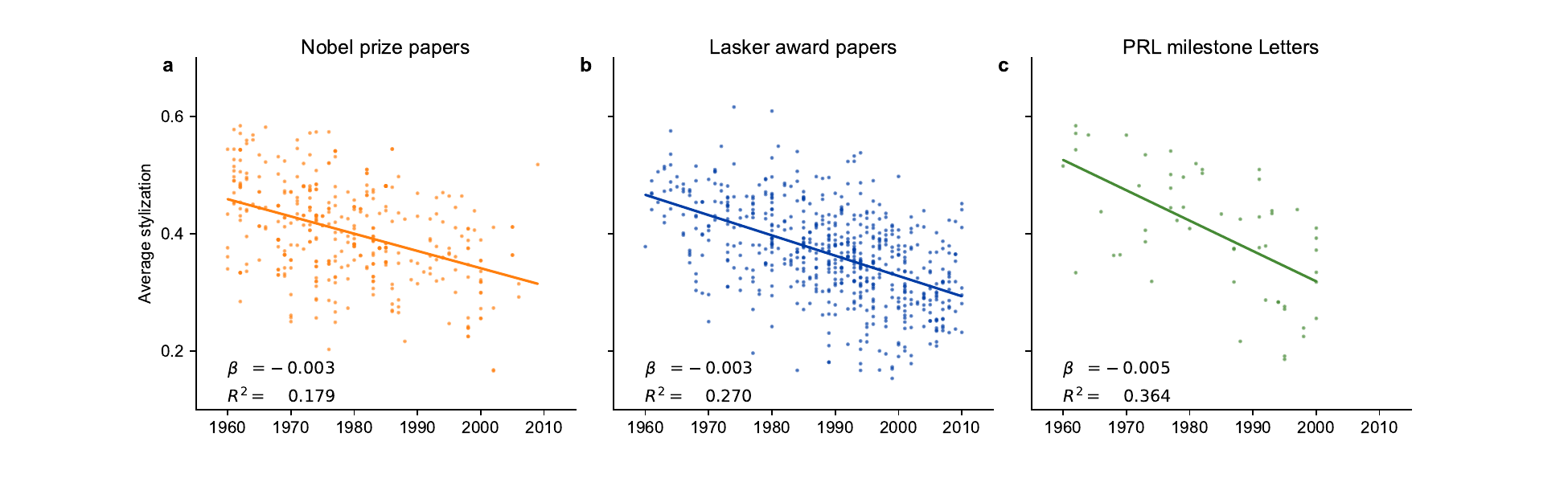}
\caption{
\textbf{Awarded papers are experiencing a decline in stylization.}
We examined 368 Nobel-winning papers (\textbf{a}) and identified a congruent pattern, consistent with the robustness check mentioned earlier. Similarly, we also analyzed 649 Lasker-winning papers (\textbf{b}) and 55 PRL-anniversary papers (\textbf{c}). PRL milestone letters experience a more pronounced decline ($\beta=-0.005$), but the overall trend is consistent with the others.
In \textbf{a}-\textbf{c}, each point here represents a papers. Additionally, in the lower-left corner, we provided the slope ($\beta$) and the coefficient of determination ($R^2$) based on linear regression estimates.
}
\label{fig:awarded_decline}
\end{figure}

As part of the robustness check, we further explored whether award-winning papers in various fields demonstrate similarly decreasing trends in Fig.~\ref{fig:awarded_decline}. To be specific, the stylization scores for winning papers, whether they received general awards like the Nobel Prize or specialized honors like the Lasker Prize and Physical Review Letters (PRL) milestone papers, have shown a decline. Among these honors, the Nobel Prize is bestowed upon scientists. Li et al.~\cite{li2019dataset} employed natural language processing techniques to pinpoint the relevant papers from the prize motivation. Our analysis revealed that papers related these awards is currently exhibiting a downward trend (see Fig.~\ref{fig:awarded_decline}a). Likewise, the Lasker Prize is also awarded to scientists; therefore, manual matching between prize winner and papers was conducted (see Fig.~\ref{fig:awarded_decline}b). Besides, as depicted in Fig.~\ref{fig:awarded_decline}c, we compiled milestone papers in the field of Physics published between 1958 and 2001, as a commemoration of the 50th anniversary of the establishment of PRL. Clearly, the stylization scores of these papers have shown a downward trend.

\begin{figure}[H]
\centering
\includegraphics[scale=0.5]{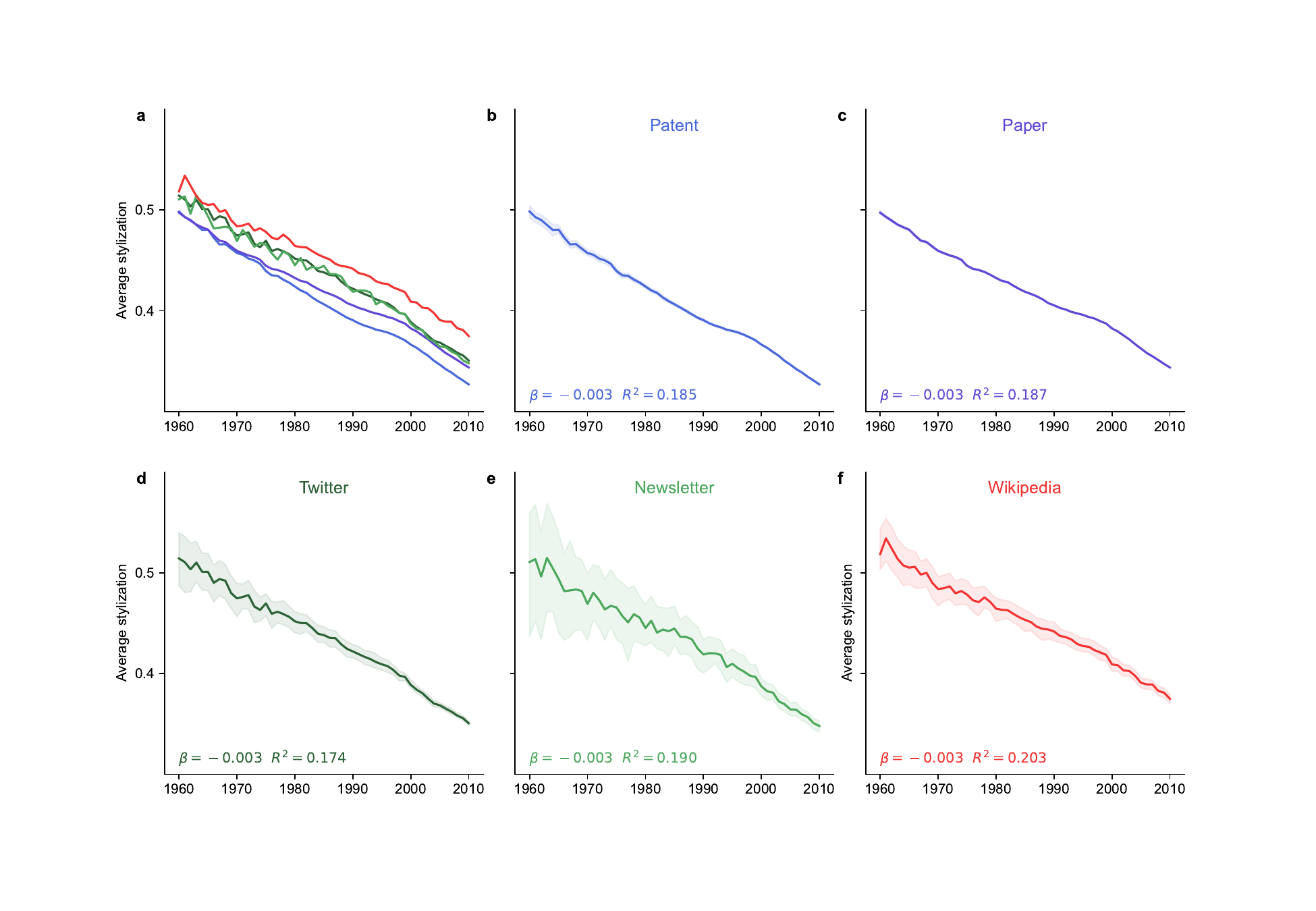}
\caption{
\textbf{Disseminated papers are also experiencing a significant decline in their stylization.}
\textbf{a}, The decline in stylization is not limited to papers disseminated within the scientific community but also extends to those cited in patents, mentioned on Twitter, reported in the media, and referenced in Wikipedia. In our analysis, we examined the datasets including 6.41 million scholarly papers cited in patents, 65.08 million referenced in scientific publications, 3.27 million discussed in Twitter posts, 0.26 million mentioned in media sources, and 0.22 million alluded to in Wikipedia articles. This phenomenon of declining stylization persists.
In (\textbf{b}-\textbf{f}), each line is presented with 95\% confidence intervals shaded around them. The slope and r-squared values, derived from linear least-squares regression estimation, are labeled in the bottom left corner. The citation links between papers and patents were obtained from the United States Patent and Trademark Office. The sources for Twitter mentions and media reports were derived from Altmetric. References cited by online encyclopedia were parsed from Wikipedia articles.
}
\label{fig:boundary_decline}
\end{figure}

Aside from the award-winning papers, are specific papers, especially those with a consistent readership, also experiencing a decline in stylization? In Fig.~\ref{fig:boundary_decline}, we compiled the papers cited by science, technology, and media outlets and organized them to illustrate the temporal trend. The empirical findings suggest that these papers, which have attracted attention outside the scientific community, are facing similar challenges. Collectively, Figs.~\ref{fig:quality_decline},~\ref{fig:awarded_decline} and~\ref{fig:boundary_decline} uncover the fact that the phenomenon extends beyond papers exclusively circulating within the scientific community; papers disseminated among other communities are also experiencing a decline.

\begin{figure}[H]
\centering
\includegraphics[scale=0.5]{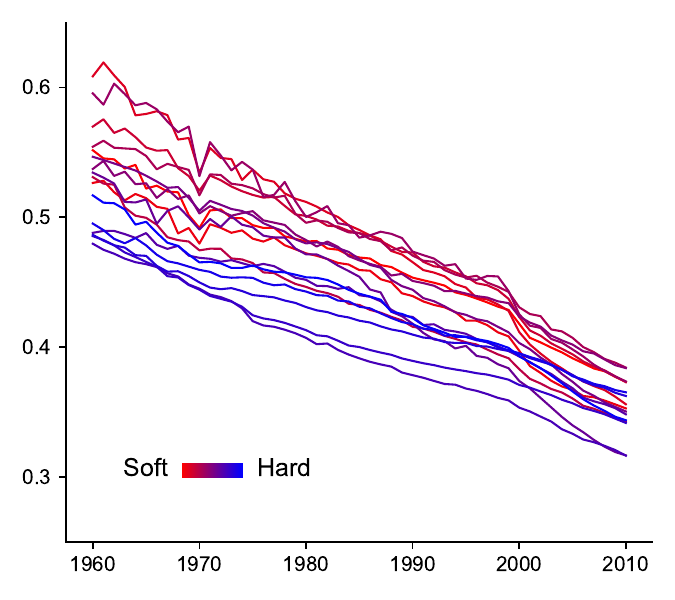}
\caption{
\textbf{The challenge of losing stylization persists from soft sciences to hard sciences.}
To investigate if this decline is limited to disciplines that heavily use mathematical and statistical tools, we constructed a continuous scientific spectrum. Specifically, we employed the social sciences as a proxy for the soft sciences and mathematics as a proxy for the hard sciences, placing each discipline on the spectrum from soft to hard sciences through neural networks. The soft sciences, as exemplified by the social sciences, exhibited a significantly higher stylization score. This could likely be attributed to the broader diversity in theories, conceptualizations, and models within the soft sciences, leading to a more stylized research environment in these disciplines.
}
\label{fig:soft_hard}
\end{figure}

In Fig.~\ref{fig:soft_hard}, we categorized 292 fields into 19 broad disciplines and excluded three disciplines (history, art, and philosophy), to examine the variations in stylization between the soft sciences and the hard sciences~\cite{peng2021neural}. Importantly, research in the soft sciences tends to exhibit more stylization, although the distinctions between soft and hard sciences have become increasingly subtle. This phenomenon can be attributed to the standardized research paradigms prevalent in fields like Mathematics and Physics, resulting in stylization similarities. In contrast, the social sciences exhibit a more extensive variety of conceptualizations, theories, and methods, which contributes to their higher level of stylization, as pointed out by~\cite{collins1994social,lamers2021meta}. Besides, the growing adoption of quantitative methods and interdisciplinary approaches may play a role in diminishing the stylization divide between the soft and hard sciences.

The extensive survey suggests an overarching pattern of declined stylization in scientific literature. Notably, even when stratifying results by the prestige of authors or the recognition received by the work, the trend persists unabated. This homogenization might be indicative of a broader cultural shift within academia towards more standardized and possibly more accessible communication. However, it also raises questions about the potential erosion of scientific stylization that were characterized, as well as the implications such a loss may bear on scientific discourse and innovation.

\clearpage
\section{Identifying the knowledge recombination}

\begin{figure}[H]
\centering
\includegraphics[scale=0.5]{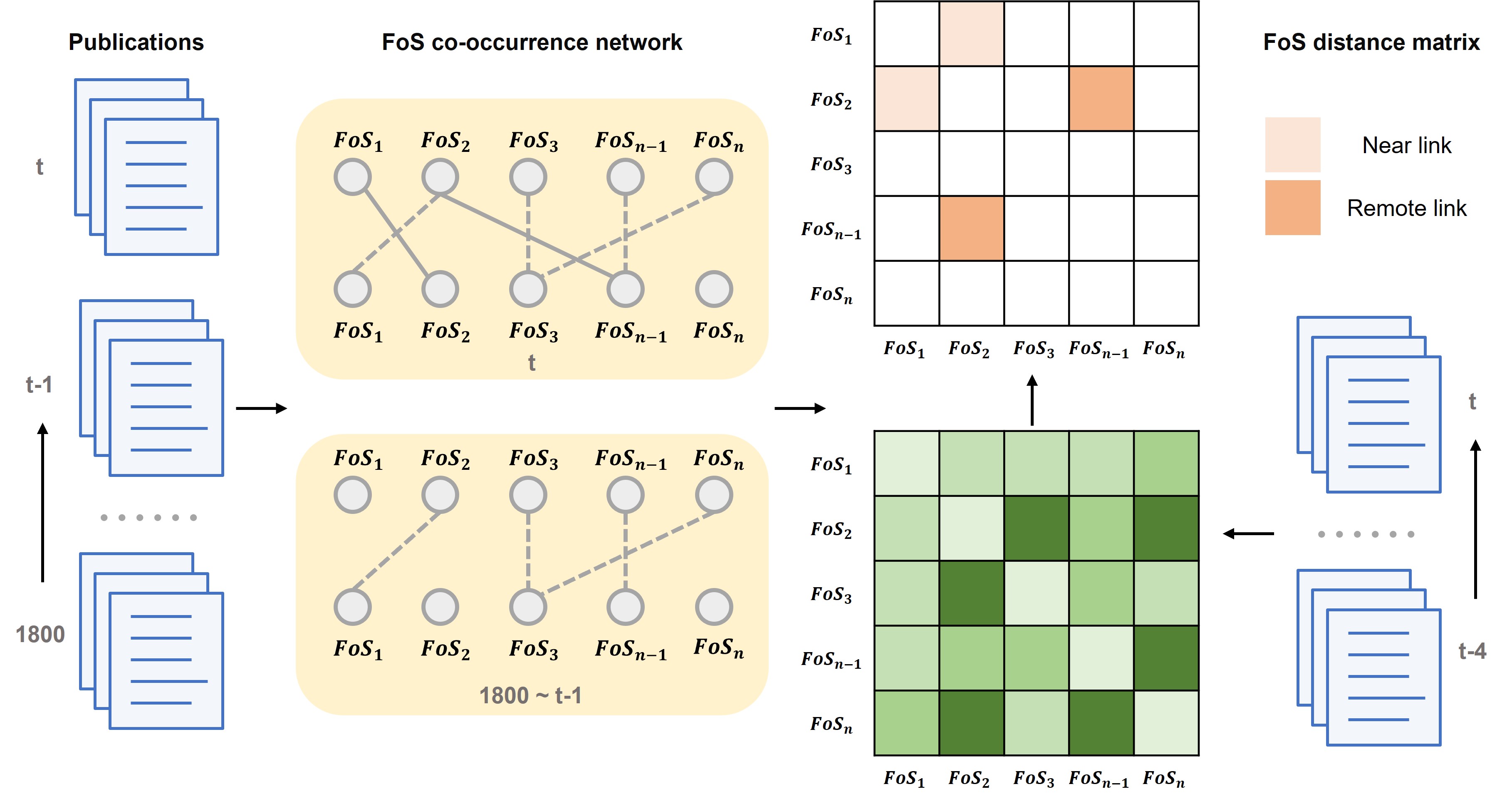}
\caption{
\textbf{Illustration of identifying the new combinations and measuring their distance.}
Grey solid links embody the emergence of new associations between knowledge components within a specific period $t$. These links signify fresh integrations within the academic discourse during this timeframe. Conversely, grey dashed links represent pre-existing recombinations, highlighting established connections that had already been formed prior to period $t$. This dichotomy provides a nuanced understanding of both the evolution and continuity within the FoS.
}
\label{fig:remote_link_example}
\end{figure}

\begin{figure}[H]
\centering
\includegraphics[scale=0.5]{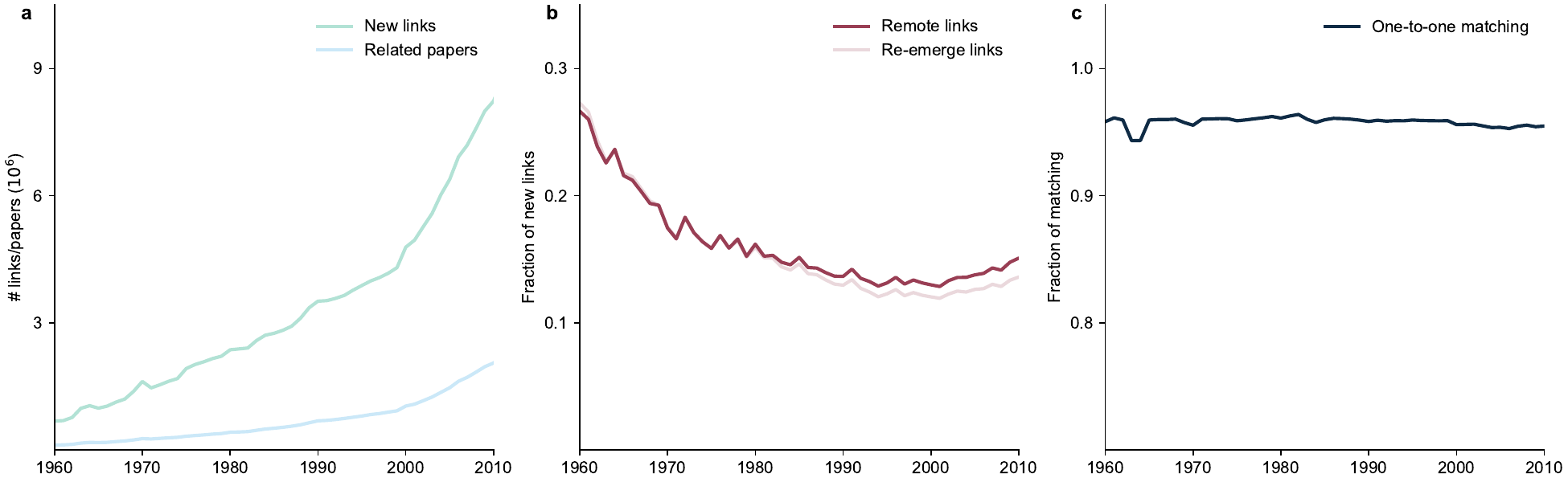}
\caption{
\textbf{New connections between concepts emerge.} %
\textbf{a}, Drawing from the annual data on the number of newly emerging knowledge combinations and the corresponding number of papers introducing these combinations, currently with prior researches, it is evident that despite the exponential growth in annual scientific publications, the quantity of newly introduced combinations follows a linear growth trend.
\textbf{b}, The proportion of distant combinations with a threshold exceeding 0.5 was approximately 27\% in 1960, gradually declining to less than 20\% thereafter, with a slight recovery around 2000.
\textbf{c}, The ratio of a single originator that has remained consistent over the past half-century. This suggests that the majority of knowledge combinations have a single point of origin, a pattern that has remained remarkably consistent. This could be attributed to the originality and novelty inherent in these research endeavors, where unique conceptual associations are often created. It underscores the pivotal role that individual papers play in introducing and propagating new combinations of knowledge throughout various FoS.
}
\label{fig:sup_know_combo}
\end{figure}

Following Schumpeter's perspective~\cite{schumpeter1934theory}, innovation entails the recombination and integration of existing knowledge. In alignment with this concept, some researchers have developed methods to quantify knowledge combinations in diverse entities, including research papers~\cite{uzzi2013atypical,foster2015tradition} and patents~\cite{fleming2001recombinant}. Inspired by these efforts, we have also constructed a similar network to observe how stylized research extends the endless frontiers of science by combining existing knowledge. Specifically, as shown in \ref{fig:remote_link_example}, we extracted the relevant concepts from each paper in the MAG database. Drawing from recent machine learning advancements, we relied on 714,971 machine-assigned labels rather than keywords provided by authors in titles or abstracts for the analysis. This approach presents two key advantages. Firstly, it obviates the necessity for operations such as lemmatization or stemming to ensure similar terminology garners identical labels, thereby sidestepping polysemy issues where the same label has diverse meanings across various domains. 
Secondly, this approach accommodates situations where novel knowledge combinations may not have been explicitly articulated by authors, or where the terminology they employed might have been interpreted differently by subsequent disseminators. To exemplify, the first case we considered involves the original patent for Bluetooth technology, which did not employ the term \textit{Bluetooth}~\cite{arts2021natural}. In a second example from scientometrics, Eugene Garfield introduced the concept of \textit{delayed recognition} in 1980~\cite{garfield1980premature} , which was latter re-conceptualized as \textit{Sleeping Beauty}~\cite{van2004sleeping} by van Raan in 2003 and subsequently gained widespread recognition~\cite{ke2015defining}. It is noteworthy that 4,150,460 papers published before 1960 were employed to construct the baseline network. This enabled us to pinpoint 9,882,021 concept pairs that have been extensively integrated by humanity since the advent of the industrial age. For that reason, we have discerned 238,028,551 paired knowledge combinations introduced since 1960. And we presented the annual distribution in Fig.~\ref{fig:sup_know_combo}a.

As noted earlier, certain knowledge recombination spans across disciplinary boundaries. To quantify the distance between these combined elements~\cite{ke2019identifying}, we reverted to the perspective of that era to scrutinize their interconnections. Specifically, for two concepts initially combined in the year $i$, we compiled all papers published from $i-4$ year to $i$ year and constructed a network to document the associations with these concepts. At that point, we conducted random walks on these annual snapshots to obtain low-dimensional representations of each node, thus gauging the distance between the concepts. 
At the same time, we distinguished distant combinations from proximate ones based on the median distance across these annual combinations. More precisely, each year, the combinations with the distance more than 0.5 are defined as linking distant concepts. Worth nothing is the fact that the percentage of distant combinations with a threshold exceeding 0.5 was around 27\% in 1960, subsequently showing a gradual decrease to less than 20\% (see Fig.~\ref{fig:sup_know_combo}b). Nonetheless, the scarcity of remote pairings is not limited to retain those links with low uptake. Even when we narrowed our analysis down to those combinations that have been reused, there was no noticeable shift in the proportion of distant connections. Recognizing that it is easier to define a widely accepted term within the same discipline than across multiple ones, we use machine-assigned labels in this study to impartially evaluate distant connections.

Simultaneously, we calculated the frequency of these combinations, referred to as ``reuse'', to measure the influence of each pair of concepts in subsequent research. This metric is analogous to citation counts and exhibits the same monotonically increasing trend.%
Additionally, we proceeded to align these concepts with the papers in which they were initially proposed. This task posed some challenges, due to the absence of clear publication dates in some early papers, it became challenging for us to precisely establish the chronological order of papers introducing the same pair of new concepts within the same year. Fortunately, while certain concepts became widely used in subsequent years, their usage in the first year was quite limited. As a result, the papers that initially introduced these concepts encountered minimal competition, as one-to-one match accounted for 95\% of the new concept combinations in the sample (see Fig.~\ref{fig:sup_know_combo}c).

\clearpage
\section{Quantifying the inconsistency on consolidation and disruption}

Disruption is an essential characteristic in measuring research innovation. Based on this premise, \cite{funk2017dynamic} introduced a novel indicator grounded in network dynamics. It serves to examine a patent's role in either reinforcing or destabilizing existing technology trajectories. This metric was subsequently validated by~\cite{wu2019large} using datasets that encompass papers sourced from Web of Science. Following these prior research, to calculate the disruption index $CD$ for a given paper denoted as $p$, published at time $t$, we considered all the papers $r_p$ cited by $p$ as well as all the papers $c_p$ citing either $p$ or $r_p$ after time $t$. Among the papers in $c_p$ that cite paper $p$, we denoted them as $f_i = 1$, whereas those in $c_p$ citing the papers in $r_i$ are labeled as $b_i = 1$. For simplicity, we calculated the disruption index ($CD$) for the focal paper $p$ using the following formula: \begin{align}CD = \frac{1}{n} \sum_{n = i}^{n} (- 2 f_{i} b_{i} + f_{i})\end{align} This indicator varies within the range of -1 to 1, elucidating the extent to which the work $i$ redirects the attention of subsequent researchers. A work with a $CD$ value exceeding 0 is interpreted as exerting a disruptive impact. It redirects the subsequent research focus away from its cited sources, compelling attention towards its own findings and insights, as elucidated by~\cite{wu2019large}. Conversely, a work with a $CD$ value less than 1 implies that it assimilates itself into the preexisting knowledge base and is subsequently employed in conjunction with it. In this study, we employed this indicator to investigate the relationship between stylized works and the established knowledge base. 

As elaborated in the main text, historic accomplishments are never wholly overlooked. This is particularly salient within the realm of interdisciplinary studies where scholars increasingly merge theories, methodologies, and resources from an array of fields to address complex challenges that single disciplines found daunting. Consequently, a simultaneous process ensues where existing knowledge is fortified, while concurrently paving the way for novel research trajectories. The ``CD'' metric can be seen as a mean statistical measure, with its demonstration reflecting the degree of fidelity to previous perspectives and foundations through the assessment of variance. In this research, we decoupled the $CD$ into $C^{\prime}$ and $D^{\prime}$ to examine the dual role of focal work in diverting attention away from the existing knowledge base and technology trajectory. 
Specifically, we approached each reference $p$ in the $r_p$ individually and examine how subsequent papers in $c_p$ are chosed between this reference $p$ and the focal work $i$. This procedure echoes the computation of $CD$, albeit each calculation assumes that $i$ has only one reference. Put plainly, for paper $i$ with $|r_p|$ references, we need to repeat this process $p$ times.
At the same time, we did not perform the subtraction operation as in the above formula. The consolidation $C_j$ is computed as: \begin{align}C_j = \frac{1}{n} \sum_{n = i}^{n} (- f_{i}^{j} b_{i}^{j})\end{align} Similarly, the destabilization $D_j$ is calculated using the formula: \begin{align}D_j = \frac{1}{n} \sum_{n = i}^{n} (- f_{i}^{j} b_{i}^{j} + f_{i}^{j})\end{align} Finally, we aggregated them to reflect their distinct stances towards established research as follows: \begin{align}C^{\prime} = \sqrt{\frac{1}{|r_p|}\sum_{j=1}^{|r_p|}(C_j - \overline{C_j})^2}\end{align} \begin{align}D^{\prime} = \sqrt{\frac{1}{|r_p|}\sum_{j=1}^{|r_p|}(D_j - \overline{D_j})^2}\end{align} We also considered measures of consistency for the original $CD$: \begin{align}{CD}^{\prime} = \sqrt{\frac{1}{|r_p|}\sum_{j=1}^{|r_p|}((D_j - C_j) - \overline{D_j - C_j})^2}\end{align} Engaging in this practice not only allows for a meticulous assessment of the varying influence of the focal paper on the existing body of literature, but also helps mitigate the disruptive effects stemming from the sheer quantity of references. Moreover, it assists in counterbalancing any potential distortions in the disruption index caused by the sheer number of references associated with the work. This ensures a more accurate representation of the individual impact of each reference on the overall disruptive capacity of the paper in question.

\clearpage
\section{Robustness check for continuous innovation}

\begin{figure}[H]
\centering
\includegraphics[scale=0.47]{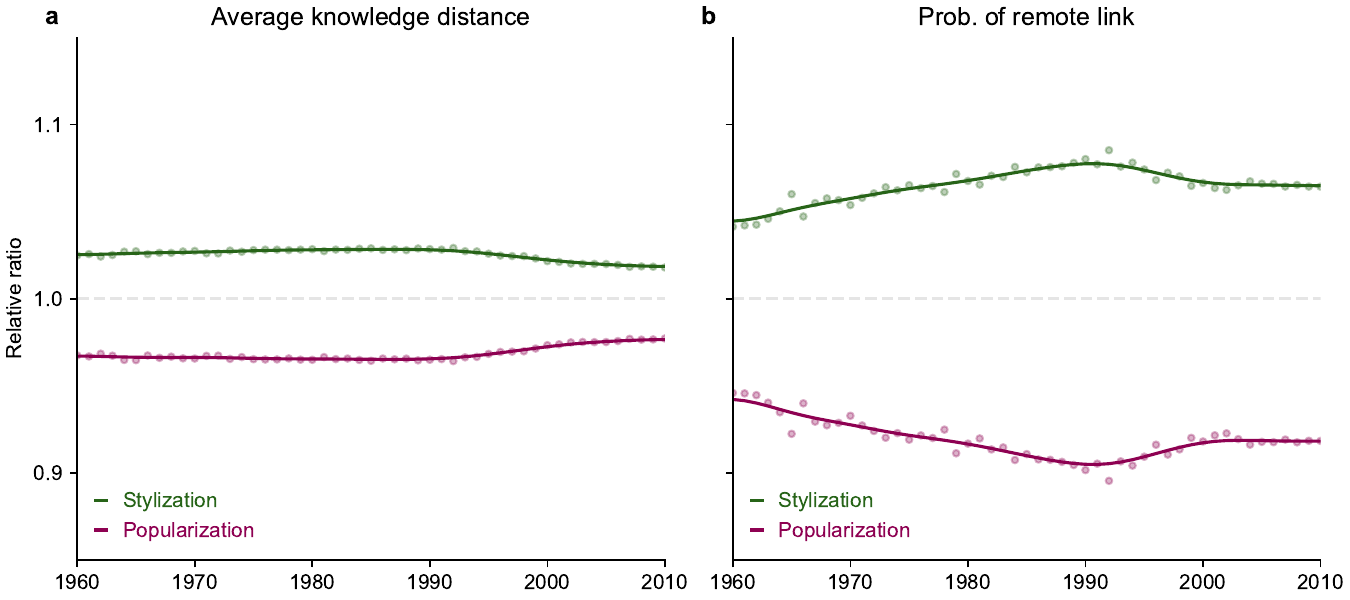}
\includegraphics[scale=0.47]{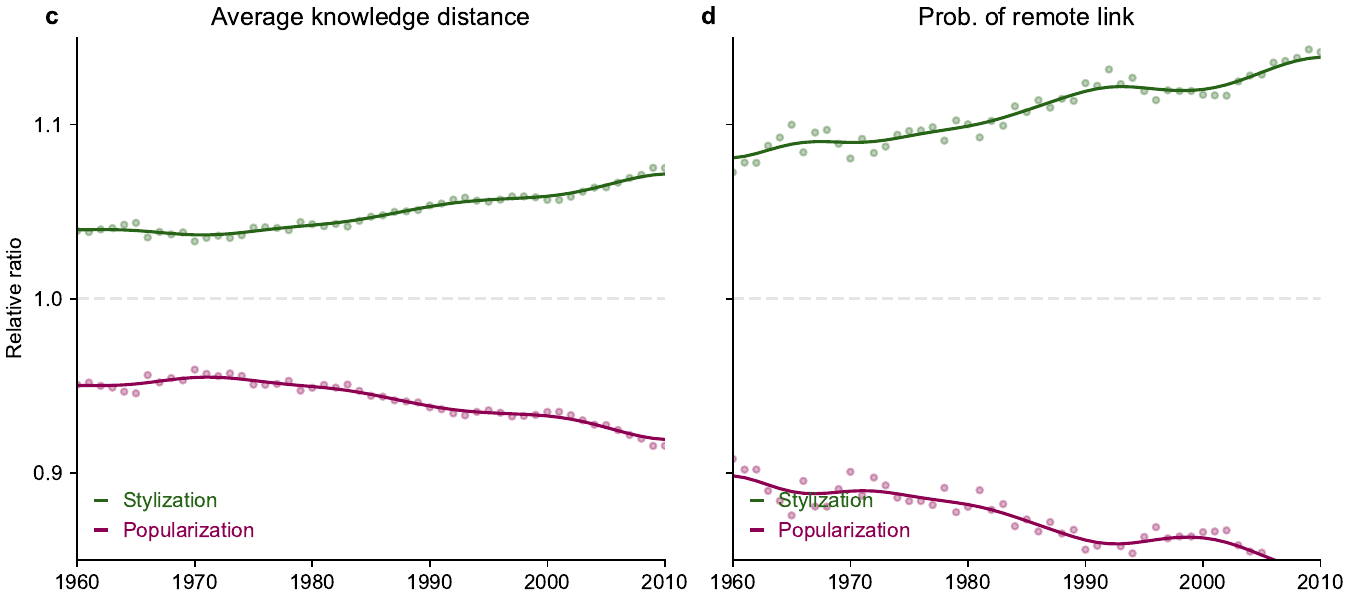}
\includegraphics[scale=0.47]{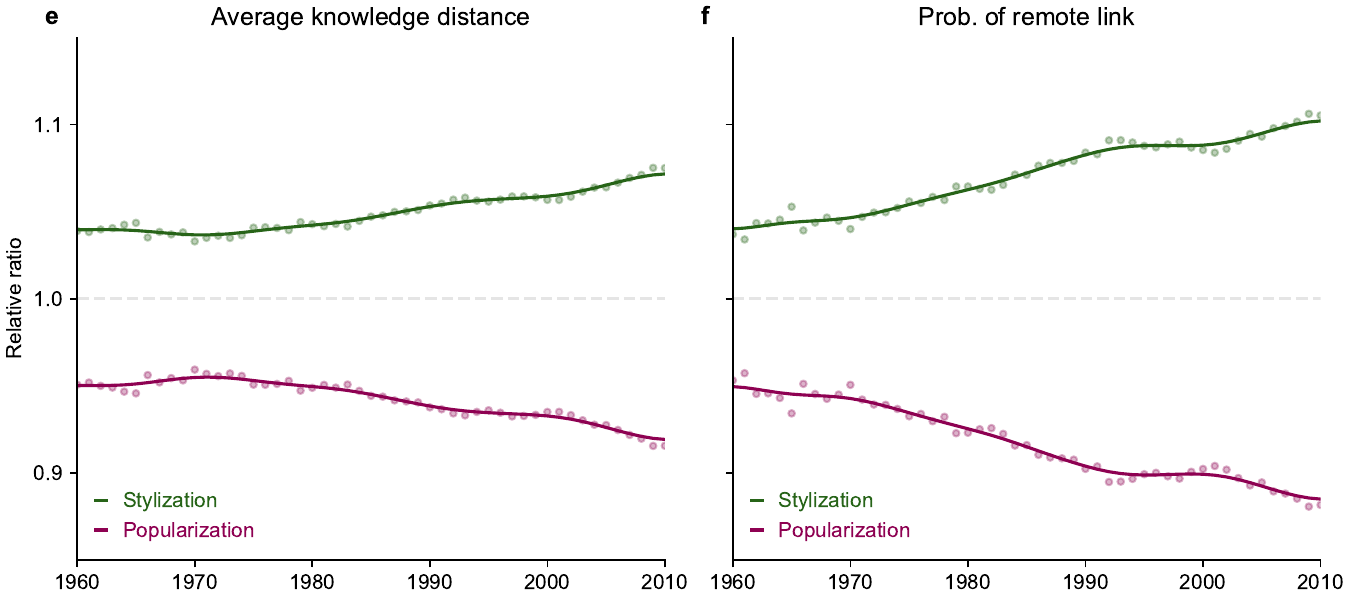}
\caption{
\textbf{Robustness check for knowledge combination.}
We excluded the papers that did not introduce novel knowledge combinations (\textbf{a}), and defined the remote links as those where the distance between associated components exceeds 0.4 (\textbf{b}) and 0.6 (\textbf{c}).
}
\label{fig:remote_robust}
\end{figure}

Two questions linger concerning how a stylized piece connects distant knowledge. The first question concerns the exclusion of papers that do not introduce novel combinations, and the second one pertains to the plausibility of this definition of distant linking. Here, we examined the first question in Fig.~\ref{fig:remote_robust}a-b, and confirmed the second one in Fig.~\ref{fig:remote_robust}c-f. 

\begin{figure}[H]
\centering
\includegraphics[scale=0.5]{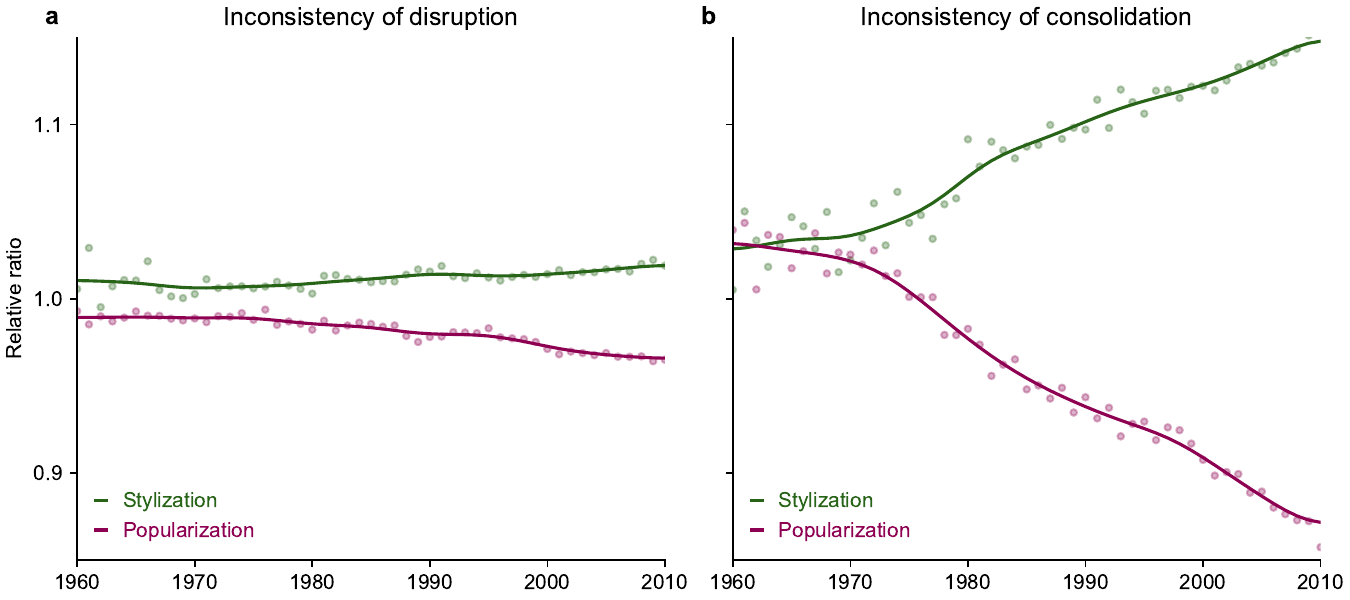}
\caption{
\textbf{The inconsistency of consolidation and disruption for highly stylization and popularization research.}
We sampled papers whose stylization score is two standard deviations above or below that of other papers from the same field in the same year and labeled them as stylized and popularized studies, respectively.
}
\label{fig:cd_pair_2std}
\end{figure}

For the $CD$ variant, we checked whether similar dual role exist in extremely stylized papers (see Fig.~\ref{fig:cd_pair_2std}).

\clearpage
\section{Robustness check for incessant bias}

In Fig.~\ref{fig:year_citation}a, we pivoted to examine the challenges stylized works encounters within the rigorous academic review process. This shift in focus brings to light a paradoxical phenomenon. While stylized research excels in pioneering new streams of knowledge, it faces an extended and intricate review period that ultimately results in a lower-than-expected citation. This paradox prompts a deeper exploration of the intricate interplay between the imaginative pursuit of stylized research and its practical challenges inherent within the scholarly community.

\begin{figure}[H]
\centering
\includegraphics[scale=0.5]{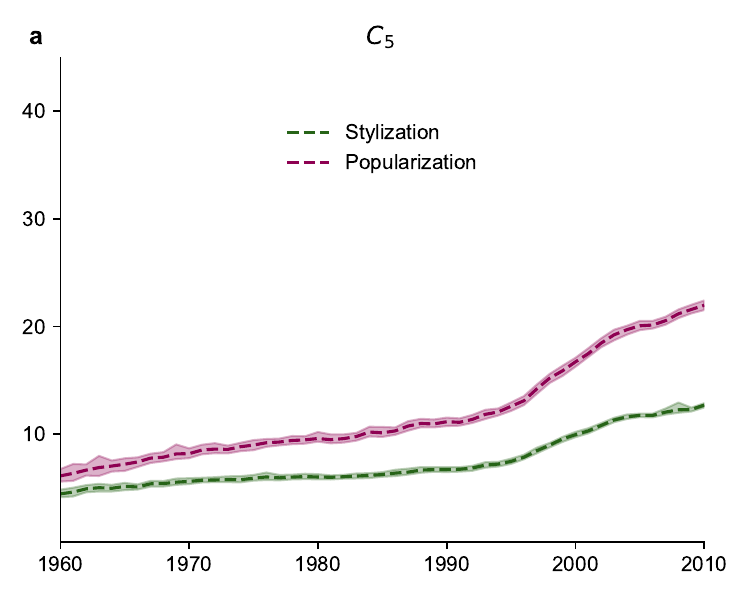}
\includegraphics[scale=0.5]{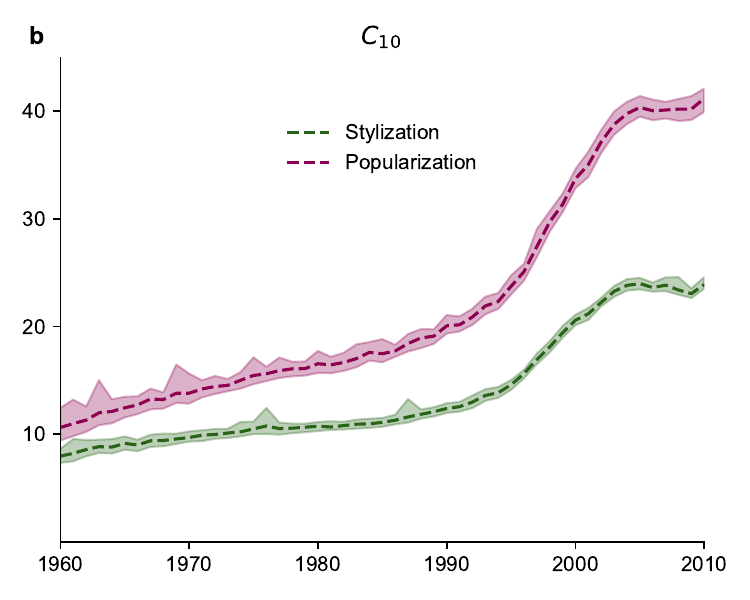}
\caption{
\textbf{Details of citations received for stylized and popularized works.}
Here, we disclose the trends of $C_5$ (\textbf{a}) and $C_{10}$ (\textbf{b}), the graphical implications of which align with the content of Fig.~\ref{fig:uscore_benefit}b-c in the main text.
}
\label{fig:c5_c10_supp}
\end{figure}

We have magnified the intricate details presented in Fig.~\ref{fig:year_citation} within the context of Fig.~\ref{fig:c5_c10_supp}. The divergence between the work characterized by stylization and that is popularized, with respect to their performance on engender citations and followers, is progressively widening. Individuals who devote themselves to endless frontier are, regrettably, less predisposed to receive higher citations. This disparity endures throughout a decade-long span; a work that gained popularity in 2010 can amass more than 40 citations over the course of a decade, whereas a stylized piece earns a modest 25 citations during the same time window, accounting for less than two-thirds of the former.
 
\begin{figure}[H]
\centering
\includegraphics[scale=0.5]{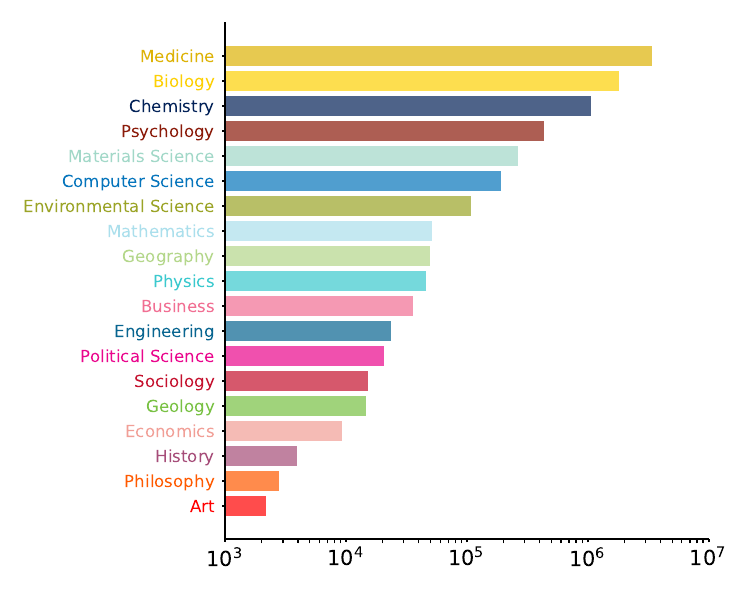}
\caption{
\textbf{Fields of papers with publication history in the PubMed database.}
We extracted the DOI of each \textit{PubMed} paper and matched it to records in \textit{MAG} to obtain field information at the level-0. In cases where papers belonged to more than one field, we applied the fraction counting method. Notably, the fields of \textit{art}, \textit{philosophy} and \textit{history} accounted for the least, while the remaining 16 fields each had over 10,000 submissions.
}
\label{fig:pubmed_coverage}
\end{figure}

\begin{figure}[H]
\centering
\includegraphics[scale=0.5]{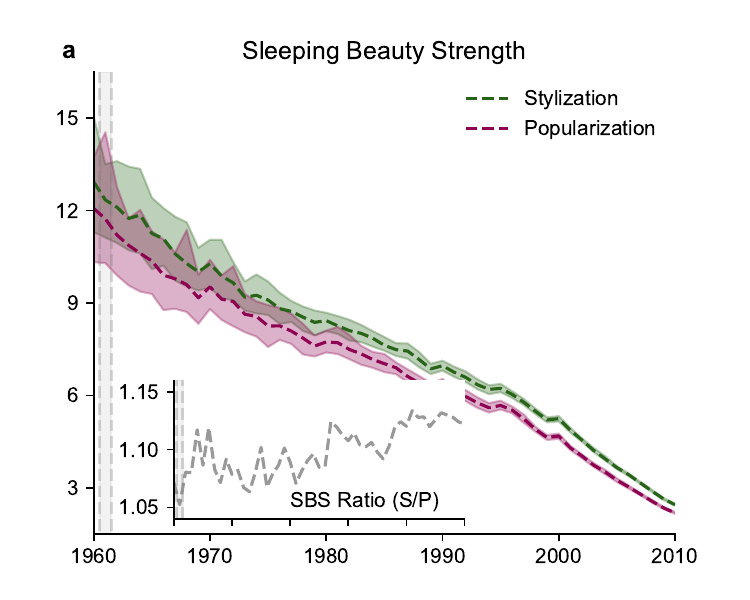}
\includegraphics[scale=0.5]{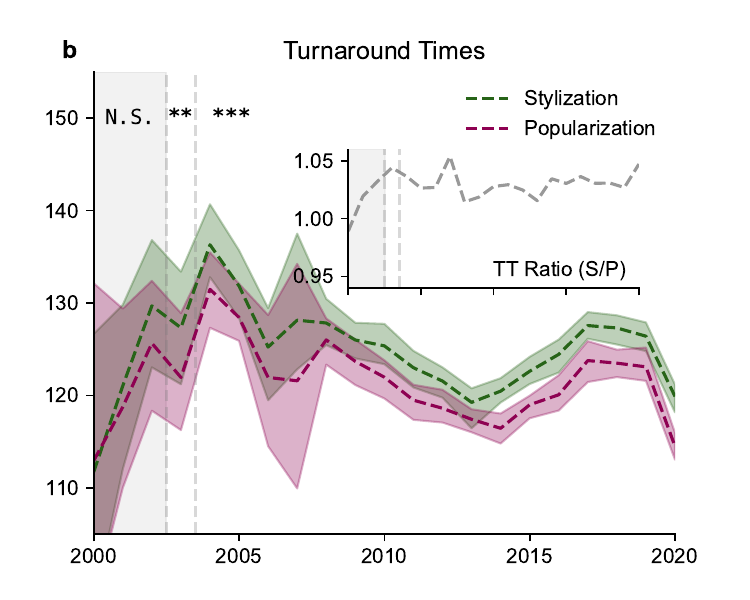}
\caption{
\textbf{Robustness check for delayed burst and turnaround times when considering outliers.}
\textbf{a}, As we extended the analysis to 2010, the phenomenon of delayed burst endured by stylized works becomes even more conspicuous. Nonetheless, it is imperative to underscore that longer observation window is imperative for a responsible assessment of sleeping beauty strength.
\textbf{b}, We also considered outliers with significantly unusual prolonged turnaround times, and the more comprehensive sample of 4,738,311 papers validated our conclusions. Stylized papers experience increasingly bias in turnaround times.
}
\label{fig:sbs_lag_supp}
\end{figure}

\begin{figure}[H]
\centering
\includegraphics[scale=0.5]{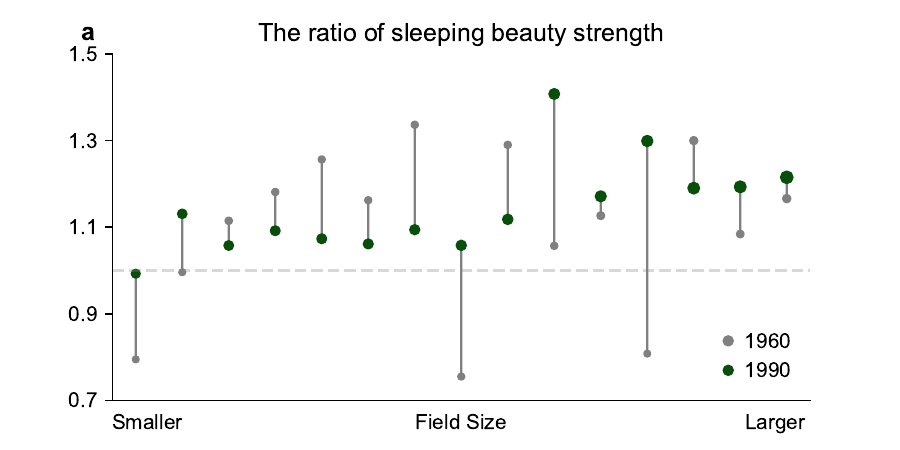}
\includegraphics[scale=0.5]{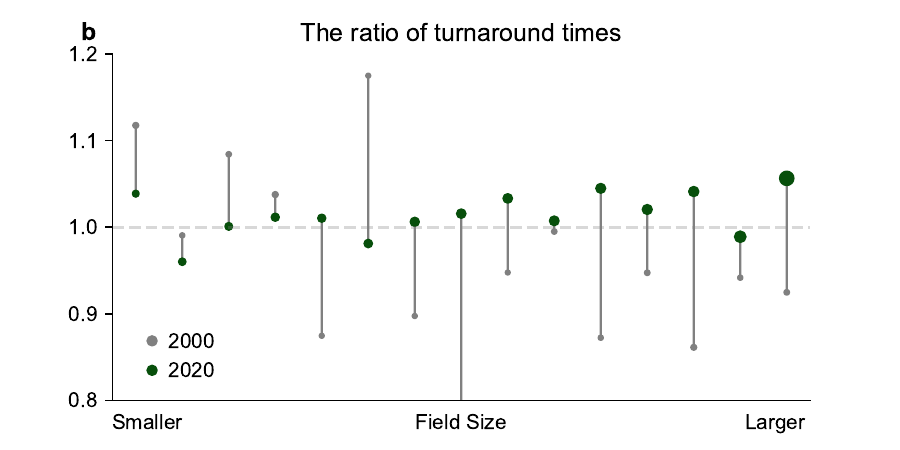}
\caption{
\textbf{Robustness check for delayed burst and turnaround times when controlling for fields.}
\textbf{a}, We selected 15 largest fields in 1960, and compared the sleeping beauty strength between stylized papers and popularized papers in 1990. Each grey point represents the ratio for a field in 1960, while green points represent the level for the same field in 1990. More than half of the fields have witnessed a more pronounced delayed burst. Most critically, areas with declining rates still fail to recognize stylized ones timely. Nearly all fields shift from timely valuation of stylized papers to prompt recognition of popularized papers, expect for medicinal chemistry. 
\textbf{b}, We employed a similar approach, selecting the 15 largest fields in 2000 and tracking the turnaround times ratio between stylized papers and popularized papers in 2020. Each grey point reflects the ratio for a field in 2000, while green points represent the level for the same fields in 2020. The size of each point indicates the scale of publications, and longer vertical lines signify more pronounced variations. The field experiencing the most substantial change is ecology, increasing from a ratio of 0.69 in 2000 to 1.02 in 2020. We sorted these fields by publication counts in 2020. Unfair treatment in review varies by discipline, but the greater the discipline, the more pronounced the issue becomes. This aligns with the findings of existing research claims~\cite{chu2021slowed}, which suggests that the impact of larger fields on the hidden dynamics of science could be profound. Among the 15 fields, 10 exhibited reduced turnaround times for stylized papers compared to 2000. The shift has been swift, with only editors and reviewers in three areas still offering faster processing for stylized works.
In \textbf{a}, \textbf{b}, larger the point means more publications. The longer vertical lines represent more variation.
}
\label{fig:field_sbs_lag}
\end{figure}

The publication history data we used is derived from the \textit{PubMed} database, which is widely recognized for covering primarily biomedical literature. However, it also includes fields such as physics. Therefore, we examined the field distribution of those papers with publication histories in Fig.~\ref{fig:pubmed_coverage}. In Fig.~\ref{fig:sbs_lag_supp}, we have broadened the dataset for robust checking for the findings related to the sleeping beauty strength and their associated turnaround time. Additionally, we delved into the differences across various fields. These disparities were discernible but not sufficient to affect our conclusions (see Fig.~\ref{fig:field_sbs_lag}).

\clearpage
\section{Regression models}\label{app:rm}
With regression analysis, this section evaluates the relation between stylization and the related feedback from scientific communities, based ordinary least squares (OLS) models and Poisson models. The OLS regression with fixed effect can be represented as follows: 

\begin{equation}
\begin{split}
y_i & = {\beta}_{s} {s}_i + {\beta}_{ts} {ts}_i \\
& + {\beta}_{c_\mu} {c_\mu}_i + {\beta}_{k_\theta} {k_\theta}_i + {\beta}_{k_\mu} {k_\mu}_i + {\beta}_{k_\theta} {k_\theta}_i \\
& + {\beta}_{rc} {rc}_i + {\beta}_{fs} {fs}_i + {\beta}_{year} {year}_i + {\beta}_{fos} {fos}_i
\end{split}
\end{equation}

This model performs well on dependent variables such as the citation normalized, sleeping beauty strength and turnaround times. Nevertheless, the excessively discrete distributions of ${C}_5$, ${C}_{10}$ and citation count notably contravene the heteroscedasticity assumped by the OLS model. As a result, this deviation leads to the inaccuracy of standard error estimated by OLS model. To address this issue, one might consider employing negative binomial regression, which is apt for count variables where variance significantly exceeding the standard deviation. Despite the widespread adoption of this model, it often struggles with convergence in the context of large-scale dataset. Therefore, to obtain accurate estimates while ensuring better convergence, we have explored Poisson regression models that employ pseudo-maximum likelihood estimation. These have been specifically tailored to ${C}_5$, ${C}_{10}$, and citation count.

\subsection*{Independent variables}

The independent variable in our model, denoted as ${s}_i$, is a continuous variable that quantifies the extent of uniqueness at the paper level. A higher ${s}_i$ indicates that paper $i$ stands further away from the mainstream research in the scientific space, representing a more stylized exploration. This can be visualized in Fig.~\ref{fig:uscore_dist}a. 
The calculation and derivation of ${s}_i$ is comprehensively outlined in the Materials and methods section. To verify the robustness of our analysis and account for potential variability, we have investigated alternative measures. These alternatives, and their implications for our findings are elaborated on in Section~\ref{app:robust_fading}. By incorporating these measures, we aimed to capture the stylization nuances across different scientific papers in a precise and reliable methodology.

\subsection*{Dependent variables}

The dependent variable in our models, $y_i$, quantifies the level of potential bias and attention experienced by a paper both before and after publication. Specifically, we examined the turnaround times from submission to publication as an indicator of pre-publication bias, and various citation metrics to quantify post-publication attention. 
To quantify the attention received by the paper from the scientific community after publication, We employed three post-publication metrics: ${C}_5$, ${C}_{10}$ and sleeping beauty strength. Also, we considered both the raw citation count and the normalized version to account for differences across fields and years. 
Furthermore, we incorporated turnaround times --- a pre-publication metric --- into our analysis as part of the dependent variable $y$ to gauge the bias faced by stylized papers within the peer-review process. This metric helps us evaluate whether more stylized works experience longer review periods.
Through the incorporation of these pre- and post-publication metrics, our analysis seeks to unravel the intricate relationship between the treatment of a paper and its adherence and deviation from prevailing trends. This in-depth analysis aims to illuminate how a paper's stylization affect the time it takes before publication and uncover the subsequent recognition it receives from scientific community.

\subsection*{Control variables}

To control the effect of potential confounding factors, we have controlled for several important variables that could potentially influence scientific production.
Firstly, team size ${ts}_i$ has been taken into account given its substantial growth over the last 20 years~\cite{wuchty2007increasing,sinatra2015century} and its significant impact on scientific production~\cite{wu2019large,jones2021rise} as noted by recent studies.

Drawing inspiration from recent research on the knowledge age~\cite{mukherjee2017nearly}, given that the search behavior concerning prior knowledge could shape the positioning of scientific outputs within the academic landscape and bring more attention, we also controlled the related factors, denoted as ${k_\mu}_i$ and ${k_\theta}_i$. It has been observed that scientific works characterized by lower mean knowledge age (${k_\mu}_i$) and higher knowledge age dispersion (${k_\theta}_i$) tend to enter the highly cited cohort.

In a similar vein, we have accounted for the mean career age ${c_\mu}_i$ and career age dispersion ${c_\theta}_i$. These factors consider the diversity of experience within research teams, where the fresh perspectives of younger scientists may complement the seasoned expertise of more established researchers~\cite{packalen2019age}. Additionally, extant researches have verified the positive correlation between reference count ${rc}_i$ and citation count (${C}_5$, ${C}_{10}$, etc.). Therefore, we included ${rc}_i$ as a control variable in our model. 

Lastly, to examine whether the observed negative effect of stylization is simply an artifact of large field~\cite{chu2021slowed}, we also considered Field size ${fs}_i$, as another control variable. By incorporating these additional controls, we aimed to paint a comprehensive and nuanced picture of the effects of stylization on the scientific production process.

\subsection*{Fixed effect}

In our model, the fixed effects are represented by ${year}_i$ for the publication year and ${fos}_i$ for the research area to which the paper belongs. The variable ${year}_i$ allows us to control for temporal variations that could affect the research outputs, such as evolving research trends, methodologies, or technology over time. Meanwhile, ${fos}_i$ accounts for unique characteristics, conventions, and citation practices inherent to various academic fields. We determine the field information based on the FoS classifications provided by the MAG. By incorporating these fixed effects into our analysis, we aim to mitigate the confounding impact of temporal and disciplinary differences, thus refining our assessment of the influence that stylization may have on a paper's reception within its scientific community.

\subsection*{Regression analysis}

Table~\ref{tab:overall} reports the outcomes of our models, reaffirming the negative correlation between stylization and citation count. The results also highlight the positive effect on sleeping beauty strength and their corresponding turnaround times. To further investigate the temporal dynamics, we have performed regression analysis across five-year intervals. We assessed $C_{5}$, $C_{10}$, citation count, citation normalized and sleeping beauty strength initiating from 1960, and analyzed turnaround times starting from 2000. The results presented in Tables~\ref{tab:c5_trend},~\ref{tab:c10_trend},~\ref{tab:cc_trend} and~\ref{tab:cn_trend} confirm that the stylized works tend to receive fewer citations. Concurrently, Table~\ref{tab:sbs_lag_trend} substantiates the escalating bias against both delayed post-publication recognition and extended pre-publication waiting periods for stylized works.

\begin{table}[H]\centering
\caption{\textbf{Regression models of stylization.}}
\label{tab:overall}
\begin{adjustbox}{max width=\textwidth}
\begin{threeparttable}
{
\def\sym#1{\ifmmode^{#1}\else\(^{#1}\)\fi}
\begin{tabular}{l*{6}{c}}
\toprule
                                             &\multicolumn{1}{c}{\shortstack{C$\textsubscript{5}$ }}&\multicolumn{1}{c}{\shortstack{C$\textsubscript{10}$ }}&\multicolumn{1}{c}{\shortstack{Citation count }}&\multicolumn{1}{c}{\shortstack{Citation normalized }}&\multicolumn{1}{c}{\shortstack{Sleeping beauty strength }}&\multicolumn{1}{c}{\shortstack{Turnaround time }}\\\cmidrule(lr){2-2}\cmidrule(lr){3-3}\cmidrule(lr){4-4}\cmidrule(lr){5-5}\cmidrule(lr){6-6}\cmidrule(lr){7-7}
                                             &\multicolumn{1}{c}{(1)}   &\multicolumn{1}{c}{(2)}   &\multicolumn{1}{c}{(3)}   &\multicolumn{1}{c}{(4)}   &\multicolumn{1}{c}{(5)}   &\multicolumn{1}{c}{(6)}   \\
\midrule
Stylization                                  &        -3.0591***&        -2.9435***&        -2.5730***&        -3.9321***&         3.5275***&        20.1456***\\
                                             &       (0.0079)   &       (0.0086)   &       (0.0114)   &       (0.0196)   &       (0.1225)   &       (0.5435)   \\
Team size                                    &         0.0020***&         0.0020***&         0.0021***&         0.0428***&        -0.0556***&         0.1077***\\
                                             &       (0.0002)   &       (0.0002)   &       (0.0002)   &       (0.0011)   &       (0.0035)   &       (0.0079)   \\
Field size                                   &        -0.0000***&        -0.0000***&         0.0000***&        -0.0000***&         0.0001***&         0.0000***\\
                                             &       (0.0000)   &       (0.0000)   &       (0.0000)   &       (0.0000)   &       (0.0000)   &       (0.0000)   \\
Reference count                              &         0.0057***&         0.0059***&         0.0061***&         0.0369***&        -0.0020***&        -0.0106***\\
                                             &       (0.0000)   &       (0.0000)   &       (0.0000)   &       (0.0002)   &       (0.0004)   &       (0.0013)   \\
Average knowledge age                        &        -0.0644***&        -0.0597***&        -0.0427***&        -0.0377***&         0.0968***&         3.7105***\\
                                             &       (0.0001)   &       (0.0001)   &       (0.0002)   &       (0.0002)   &       (0.0017)   &       (0.0183)   \\
Average career experience                    &         0.0106***&         0.0108***&         0.0126***&         0.0300***&         0.0526***&        -0.3954***\\
                                             &       (0.0001)   &       (0.0001)   &       (0.0001)   &       (0.0003)   &       (0.0018)   &       (0.0068)   \\
Knowledge age dispersion                     &         0.0261***&         0.0294***&         0.0269***&         0.0274***&         0.1265***&        -2.4682***\\
                                             &       (0.0001)   &       (0.0001)   &       (0.0002)   &       (0.0003)   &       (0.0018)   &       (0.0168)   \\
Career experience dispersion                 &         0.0089***&         0.0086***&         0.0074***&         0.0145***&        -0.0189***&         0.0889***\\
                                             &       (0.0001)   &       (0.0001)   &       (0.0001)   &       (0.0003)   &       (0.0017)   &       (0.0073)   \\
Year FE                                      &            Yes   &            Yes   &            Yes   &            Yes   &            Yes   &            Yes   \\
Field FE                                     &            Yes   &            Yes   &            Yes   &            Yes   &            Yes   &            Yes   \\
\midrule
{\( N \)}                                    &       25182058   &       25182058   &       25182058   &       25182058   &        5750318   &        5751006   \\
{(Pseudo) \( R^2 \)}                         &           0.25   &           0.24   &           0.15   &           0.03   &           0.03   &           0.05   \\
\bottomrule
\end{tabular}
}

\begin{tablenotes}
\item Models 1, 2 and 3 are deduced through Poisson regressions with pseudo maximum likelihood, while models 4, 5 and 6 are estimated by OLS regression. The robust standard errors associated with each model are represented in parentheses. 
\item {+}p$<$0.1 {*}p$<$0.05 {**}p$<$0.01; {***}p$<$0.001.
\end{tablenotes}
\end{threeparttable}
\end{adjustbox}
\end{table}

\begin{table}[H]\centering
\caption{\textbf{Regression models of $C_{5}$ and stylization over time.}}
\label{tab:c5_trend}
\begin{adjustbox}{max width=\textwidth}
\begin{threeparttable}
{
\def\sym#1{\ifmmode^{#1}\else\(^{#1}\)\fi}
\begin{tabular}{l*{10}{c}}
\toprule
                                             &\multicolumn{10}{c}{\shortstack{C$\textsubscript{5}$ }}                                                                                                                                  \\\cmidrule(lr){2-11}
                                             &\multicolumn{1}{c}{(1)}   &\multicolumn{1}{c}{(2)}   &\multicolumn{1}{c}{(3)}   &\multicolumn{1}{c}{(4)}   &\multicolumn{1}{c}{(5)}   &\multicolumn{1}{c}{(6)}   &\multicolumn{1}{c}{(7)}   &\multicolumn{1}{c}{(8)}   &\multicolumn{1}{c}{(9)}   &\multicolumn{1}{c}{(10)}   \\
\midrule
Stylization                                  &        -1.7822***&        -2.2260***&        -2.5373***&        -2.6953***&        -2.9050***&        -2.9595***&        -2.9245***&        -2.9953***&        -2.9045***&        -3.0034***\\
                                             &       (0.0500)   &       (0.0347)   &       (0.0266)   &       (0.0241)   &       (0.0240)   &       (0.0245)   &       (0.0197)   &       (0.0181)   &       (0.0152)   &       (0.0157)   \\
Controls\tnote{†}                            &            Yes   &            Yes   &            Yes   &            Yes   &            Yes   &            Yes   &            Yes   &            Yes   &            Yes   &            Yes   \\
Year FE                                      &            Yes   &            Yes   &            Yes   &            Yes   &            Yes   &            Yes   &            Yes   &            Yes   &            Yes   &            Yes   \\
Field FE                                     &            Yes   &            Yes   &            Yes   &            Yes   &            Yes   &            Yes   &            Yes   &            Yes   &            Yes   &            Yes   \\
\midrule
{\( N \)}                                    &         267721   &         486822   &         785611   &        1100054   &        1541726   &        2112906   &        2757753   &        3652780   &        5033798   &        7442878   \\
{Pseudo \( R^2 \)}                           &           0.14   &           0.16   &           0.18   &           0.18   &           0.19   &           0.20   &           0.22   &           0.24   &           0.24   &           0.23   \\
\bottomrule
\end{tabular}
}

\begin{tablenotes}
\item The estimates are derived from Poisson regressions that employ pseudo maximum likelihood. The robust standard errors associated with each model are represented in parentheses.
\item {+}p$<$0.1 {*}p$<$0.05 {**}p$<$0.01; {***}p$<$0.001.
\item[†] Controls here mean related control variables above.
\end{tablenotes}
\end{threeparttable}
\end{adjustbox}
\end{table}

\begin{table}[H]\centering
\caption{\textbf{Regression models of $C_{10}$ and stylization over time.}}
\label{tab:c10_trend}
\begin{adjustbox}{max width=\textwidth}
\begin{threeparttable}
{
\def\sym#1{\ifmmode^{#1}\else\(^{#1}\)\fi}
\begin{tabular}{l*{10}{c}}
\toprule
                                             &\multicolumn{10}{c}{\shortstack{C$\textsubscript{10}$ }}                                                                                                                                 \\\cmidrule(lr){2-11}
                                             &\multicolumn{1}{c}{(1)}   &\multicolumn{1}{c}{(2)}   &\multicolumn{1}{c}{(3)}   &\multicolumn{1}{c}{(4)}   &\multicolumn{1}{c}{(5)}   &\multicolumn{1}{c}{(6)}   &\multicolumn{1}{c}{(7)}   &\multicolumn{1}{c}{(8)}   &\multicolumn{1}{c}{(9)}   &\multicolumn{1}{c}{(10)}   \\
\midrule
Stylization                                  &        -1.6238***&        -2.0603***&        -2.3402***&        -2.4700***&        -2.7136***&        -2.7484***&        -2.7629***&        -2.8141***&        -2.8164***&        -2.9633***\\
                                             &       (0.0675)   &       (0.0414)   &       (0.0296)   &       (0.0321)   &       (0.0259)   &       (0.0296)   &       (0.0221)   &       (0.0196)   &       (0.0168)   &       (0.0167)   \\
Controls\tnote{†}                            &            Yes   &            Yes   &            Yes   &            Yes   &            Yes   &            Yes   &            Yes   &            Yes   &            Yes   &            Yes   \\
Year FE                                      &            Yes   &            Yes   &            Yes   &            Yes   &            Yes   &            Yes   &            Yes   &            Yes   &            Yes   &            Yes   \\
Field FE                                     &            Yes   &            Yes   &            Yes   &            Yes   &            Yes   &            Yes   &            Yes   &            Yes   &            Yes   &            Yes   \\
\midrule
{\( N \)}                                    &         267727   &         486822   &         785611   &        1100054   &        1541726   &        2112906   &        2757753   &        3652780   &        5033798   &        7442878   \\
{Pseudo \( R^2 \)}                           &           0.14   &           0.16   &           0.17   &           0.17   &           0.18   &           0.19   &           0.21   &           0.21   &           0.21   &           0.21   \\
\bottomrule
\end{tabular}
}

\begin{tablenotes}
\item The estimates are derived from Poisson regressions with pseudo maximum likelihood. The robust standard errors associated with each model are represented in parentheses. 
\item {+}p$<$0.1 {*}p$<$0.05 {**}p$<$0.01; {***}p$<$0.001.
\item[†] Controls here refer to the aforementioned control variables in our study.
\end{tablenotes}
\end{threeparttable}
\end{adjustbox}
\end{table}

\begin{table}[H]\centering
\caption{\textbf{Regression models of citation count and stylization over time.}}
\label{tab:cc_trend}
\begin{adjustbox}{max width=\textwidth}
\begin{threeparttable}
{
\def\sym#1{\ifmmode^{#1}\else\(^{#1}\)\fi}
\begin{tabular}{l*{10}{c}}
\toprule
                                             &\multicolumn{10}{c}{\shortstack{Citation count }}                                                                                                                                  \\\cmidrule(lr){2-11}
                                             &\multicolumn{1}{c}{(1)}   &\multicolumn{1}{c}{(2)}   &\multicolumn{1}{c}{(3)}   &\multicolumn{1}{c}{(4)}   &\multicolumn{1}{c}{(5)}   &\multicolumn{1}{c}{(6)}   &\multicolumn{1}{c}{(7)}   &\multicolumn{1}{c}{(8)}   &\multicolumn{1}{c}{(9)}   &\multicolumn{1}{c}{(10)}   \\
\midrule
Stylization                                  &        -0.6291***&        -1.3587***&        -1.5882***&        -1.7313***&        -2.0790***&        -2.2418***&        -2.3951***&        -2.5437***&        -2.7068***&        -2.9291***\\
                                             &       (0.1794)   &       (0.0875)   &       (0.1256)   &       (0.1396)   &       (0.0396)   &       (0.0447)   &       (0.0335)   &       (0.0308)   &       (0.0197)   &       (0.0173)   \\
Controls\tnote{†}                            &            Yes   &            Yes   &            Yes   &            Yes   &            Yes   &            Yes   &            Yes   &            Yes   &            Yes   &            Yes   \\
Year FE                                      &            Yes   &            Yes   &            Yes   &            Yes   &            Yes   &            Yes   &            Yes   &            Yes   &            Yes   &            Yes   \\
Field FE                                     &            Yes   &            Yes   &            Yes   &            Yes   &            Yes   &            Yes   &            Yes   &            Yes   &            Yes   &            Yes   \\
\midrule
{\( N \)}                                    &         267727   &         486822   &         785611   &        1100054   &        1541726   &        2112906   &        2757753   &        3652780   &        5033798   &        7442878   \\
{Pseudo \( R^2 \)}                           &           0.09   &           0.08   &           0.08   &           0.08   &           0.10   &           0.12   &           0.14   &           0.15   &           0.18   &           0.19   \\
\bottomrule
\end{tabular}
}

\begin{tablenotes}
\item The estimates are derived from Poisson regressions with pseudo maximum likelihood. The robust standard errors associated with each model are represented in parentheses.
\item {+}p$<$0.1 {*}p$<$0.05 {**}p$<$0.01; {***}p$<$0.001.
\item[†] Controls here refer to the aforementioned control variables in our study.
\end{tablenotes}
\end{threeparttable}
\end{adjustbox}
\end{table}

\begin{table}[H]\centering
\caption{\textbf{Regression models of citation normalized and stylization over time.}}
\label{tab:cn_trend}
\begin{adjustbox}{max width=\textwidth}
\begin{threeparttable}
{
\def\sym#1{\ifmmode^{#1}\else\(^{#1}\)\fi}
\begin{tabular}{l*{10}{c}}
\toprule
                                             &\multicolumn{10}{c}{\shortstack{Citation normalized }}                                                                                                                                  \\\cmidrule(lr){2-11}
                                             &\multicolumn{1}{c}{(1)}   &\multicolumn{1}{c}{(2)}   &\multicolumn{1}{c}{(3)}   &\multicolumn{1}{c}{(4)}   &\multicolumn{1}{c}{(5)}   &\multicolumn{1}{c}{(6)}   &\multicolumn{1}{c}{(7)}   &\multicolumn{1}{c}{(8)}   &\multicolumn{1}{c}{(9)}   &\multicolumn{1}{c}{(10)}   \\
\midrule
Stylization                                  &        -1.0975***&        -2.1150***&        -2.0803***&        -2.4130***&        -2.6073***&        -2.7642***&        -3.0648***&        -3.4852***&        -4.4408***&        -5.2349***\\
                                             &       (0.2001)   &       (0.1386)   &       (0.1169)   &       (0.1369)   &       (0.0587)   &       (0.0619)   &       (0.0513)   &       (0.0506)   &       (0.0408)   &       (0.0381)   \\
Controls\tnote{†}                            &            Yes   &            Yes   &            Yes   &            Yes   &            Yes   &            Yes   &            Yes   &            Yes   &            Yes   &            Yes   \\
Year FE                                      &            Yes   &            Yes   &            Yes   &            Yes   &            Yes   &            Yes   &            Yes   &            Yes   &            Yes   &            Yes   \\
Field FE                                     &            Yes   &            Yes   &            Yes   &            Yes   &            Yes   &            Yes   &            Yes   &            Yes   &            Yes   &            Yes   \\
\midrule
{\( N \)}                                    &         267727   &         486822   &         785611   &        1100054   &        1541726   &        2112906   &        2757753   &        3652780   &        5033798   &        7442878   \\
{Pseudo \( R^2 \)}                           &           0.02   &           0.03   &           0.01   &           0.01   &           0.03   &           0.02   &           0.03   &           0.02   &           0.05   &           0.05   \\
\bottomrule
\end{tabular}
}

\begin{tablenotes}
\item The estimates are derived from OLS regressions. The robust standard errors associated with each model are represented in parentheses. 
\item {+}p$<$0.1 {*}p$<$0.05 {**}p$<$0.01; {***}p$<$0.001.
\item[†] Controls here refer to the aforementioned control variables in our study.
\end{tablenotes}
\end{threeparttable}
\end{adjustbox}
\end{table}

\begin{table}[H]\centering
\caption{\textbf{Regression models of sleeping beauty strength, turnaround time and stylization over time.}}
\label{tab:sbs_lag_trend}
\begin{adjustbox}{max width=\textwidth}
\begin{threeparttable}
{
\def\sym#1{\ifmmode^{#1}\else\(^{#1}\)\fi}
\begin{tabular}{l*{10}{c}}
\toprule
                                             &\multicolumn{6}{c}{\shortstack{Sleeping beauty strength}}                                                    &\multicolumn{4}{c}{\shortstack{Turnaround time}}                       \\\cmidrule(lr){2-7}\cmidrule(lr){8-11}
                                             &\multicolumn{1}{c}{(1)}   &\multicolumn{1}{c}{(2)}   &\multicolumn{1}{c}{(3)}   &\multicolumn{1}{c}{(4)}   &\multicolumn{1}{c}{(5)}   &\multicolumn{1}{c}{(6)}   &\multicolumn{1}{c}{(7)}   &\multicolumn{1}{c}{(8)}   &\multicolumn{1}{c}{(9)}   &\multicolumn{1}{c}{(10)}   \\
\midrule
Stylization                                  &         3.9755***&         3.6226***&         3.4403***&         3.0889***&         3.2005***&         3.6896***&         7.6294*  &        12.1189***&        16.0384***&        22.2636***\\
                                             &       (0.9918)   &       (0.6998)   &       (0.4239)   &       (0.3134)   &       (0.1959)   &       (0.1341)   &       (3.5571)   &       (1.4587)   &       (0.9415)   &       (0.7633)   \\
Controls\tnote{†}                            &            Yes   &            Yes   &            Yes   &            Yes   &            Yes   &            Yes   &            Yes   &            Yes   &            Yes   &            Yes   \\
Year FE                                      &            Yes   &            Yes   &            Yes   &            Yes   &            Yes   &            Yes   &            Yes   &            Yes   &            Yes   &            Yes   \\
Field FE                                     &            Yes   &            Yes   &            Yes   &            Yes   &            Yes   &            Yes   &            Yes   &            Yes   &            Yes   &            Yes   \\
\midrule
{\( N \)}                                    &         240677   &         441730   &         713993   &        1005463   &        1414898   &        1933555   &         147877   &         933183   &        1897867   &        2772055   \\
{\( R^2 \)}                                  &           0.02   &           0.02   &           0.02   &           0.03   &           0.03   &           0.03   &           0.08   &           0.08   &           0.06   &           0.05   \\
\bottomrule
\end{tabular}
}

\begin{tablenotes}
\item The estimates are derived from OLS regressions. The robust standard errors associated with each model are represented in parentheses. 
\item {+}p$<$0.1 {*}p$<$0.05 {**}p$<$0.01; {***}p$<$0.001.
\item[†] Controls here refer to the aforementioned control variables in our study.
\end{tablenotes}
\end{threeparttable}
\end{adjustbox}
\end{table}

\clearpage
\printbibliography[heading=subbibliography]
\end{refsection}
\end{document}